\documentclass[manuscript]{emulateapj}

\slugcomment{Accepted by ApJ}
\shorttitle{ChanPlaNS CSPNe X-ray Emission}
\shortauthors{Montez et al.}

\begin{document}

\title{The Chandra Planetary Nebula Survey (ChanPlaNS): III. X-ray Emission from the Central Stars of Planetary Nebulae}

\author{
R. Montez Jr.\altaffilmark{1},
J. H. Kastner\altaffilmark{2},
B. Balick\altaffilmark{3},
E.  Behar\altaffilmark{4},
E.  Blackman\altaffilmark{5},
V. Bujarrabal\altaffilmark{6},
Y.-H. Chu\altaffilmark{7},
R. L. M. Corradi\altaffilmark{8,9},
O. De Marco\altaffilmark{10},
A. Frank\altaffilmark{5},
M. Freeman\altaffilmark{2},
D. J. Frew\altaffilmark{10},
M. A. Guerrero\altaffilmark{11},
D.   Jones\altaffilmark{12},
J. A. Lopez\altaffilmark{13},
B. Miszalski\altaffilmark{14,15},
J. Nordhaus\altaffilmark{16},
Q. A. Parker\altaffilmark{10,17},
R. Sahai\altaffilmark{18},
C. Sandin\altaffilmark{19},
D. Schonberner\altaffilmark{19},
N.  Soker\altaffilmark{5},
J. L. Sokoloski\altaffilmark{20},
M. Steffen\altaffilmark{19},
J. A. Toal\'a\altaffilmark{11},
T. Ueta\altaffilmark{21},
E.  Villaver\altaffilmark{22},
A. Zijlstra\altaffilmark{23} \\
(See next page for Author Affiliations)
}

\begin{abstract}
We present X-ray spectral analysis of 20 point-like X-ray sources detected in Chandra Planetary Nebula Survey (ChanPlaNS) observations of 59 planetary nebulae (PNe) in the solar neighborhood. 
Most of these 20 detections are associated with luminous central stars within relatively young, compact nebulae. 
The vast majority of these point-like X-ray-emitting sources at PN cores display relatively ``hard'' ($\geq0.5$~keV) X-ray emission components that are unlikely to be due to photospheric emission from the hot central stars (CSPN). 
Instead, we demonstrate that these sources are well modeled by optically-thin thermal plasmas. 
From the plasma properties, we identify two classes of CSPN X-ray emission: (1) high-temperature plasmas with X-ray luminosities, $L_{\rm X}$, that appear uncorrelated with the CSPN bolometric luminosity, $L_{\rm bol}$; and (2) lower-temperature plasmas with $L_{\rm X}/L_{\rm bol}\sim10^{-7}$.
We suggest these two classes correspond to the physical processes of magnetically active binary companions and self-shocking stellar winds, respectively. 
In many cases this conclusion is supported by corroborative multiwavelength evidence for the wind and binary properties of the PN central stars. 
By thus honing in on the origins of X-ray emission from PN central stars, we enhance the ability of CSPN X-ray sources to constrain models of PN shaping that invoke wind interactions and binarity. 
\end{abstract}

\keywords{planetary nebula: general,stars: evolution,stars: winds, X-rays: stars}

\section{Introduction} 

The late stages of evolution of low- to intermediate-mass stars (1-8 $M_{\sun}$) are characterized by high rates of mass loss on the red and asymptotic giant branches (RG and AGB, respectively). 
During the post-AGB phase the lost mass is shaped by a fast, possibly collimated, wind into a planetary nebula (PN) that is ionized by the bright ($10^{3}-10^{4} L_{\sun}$) and hot ($>30$~kK) central star (CSPN). 
After reaching their highest temperatures, the CSPNe turn back in the H-R diagram towards lower temperatures and luminosities as they approach the white dwarf (WD) cooling track \citep[e.g.,][]{1995A&A...299..755B}. 
This post-AGB/pre-WD evolutionary period is short-lived ($10^{3}-10^{5}$ years depending on the initial mass of the star). 
If one ignores the potentially important effects of binarity, the evolution of the PN proceeds in lock step with the CSPN on a given evolutionary mass track \citep{2004A&A...414..993P}, with the nebular density being the highest early in the evolution and decreasing as the CSPN approaches the WD cooling track. 
Peculiar spectral types are littered along this evolutionary path \citep{1991IAUS..145..375M};  some central stars show strong Wolf-Rayet-like emission lines, some share characteristics with the class of H-deficient PG1159-type stars, and others are classified as hot sub-dwarfs, sometimes in close binary systems with late-type companions.
Whether any of these peculiar spectral types follow a common evolutionary sequence is a matter of open debate \citep{2002PASP..114..602D,2006PASP..118..183W}.

PNe are commonly detected as X-ray sources \citep[Kastner et al. 2012; hereafter,][]{2012AJ....144...58K}. 
Typically, this X-ray emission from PNe comes in two forms: (1) compact point-like sources in the vicinity of the central stars and (2) extended diffuse X-ray emission from adiabatic shocks and hot bubbles \citep[see][and references therein]{2012AJ....144...58K}. 
In the 1990's, ROSAT established that most PNe detected in X-rays generated X-ray photons with energies $<0.5$~keV, consistent with emission from the photospheres of hot CSPNe \citep{2000ApJS..129..295G}. 
But a few ROSAT spectra of PNe featured X-ray emission $>$0.5 keV. 
Several of these were later established by the Chandra X-ray Observatory to originate from diffuse nebular emission \citep[e.g.,][]{2000ApJ...545L..57K}. 
The notable exception is the central star of the Helix Nebula, whose ``hard'' X-ray emission remains point-like even at Chandra's subarcsecond resolution \citep{2001ApJ...553L..55G}.
Since CSPNe can reach temperatures of 100-200 kK, photospheric X-ray emission from CSPNe is not unexpected, but any ``hard'' X-ray emission, i.e. X-ray photons with energies $>$0.5 keV, might be expected to fall below the detection threshold (see Appendix~\ref{appendixA}).
Nonetheless, energetic photons $>$0.5 keV are detected from roughly a third of those observed CSPNe (Paper I; Freeman et al. 2014, hereafter Paper II).

The origins and emission mechanisms underlying these point-like hard X-ray sources are unclear, but a few scenarios have been proposed. 
When similarly hard X-ray emitting sources are observed in white dwarfs, they are typically attributed to known binary companions with active coronae \citep{2003AJ....125.2239O,2004AJ....127..477C,2010AJ....140.1433B}. 
\citet{2010ApJ...721.1820M} argued in favor of a binary origin for hard X-ray emission from CSPNe known to harbor companions. 
In two post-common envelope binary (PCEB) central stars, DS 1 and HFG 1, and a binary CSPN with a rapidly rotating giant companion, LoTr 5, the X-ray spectra indicated plasma temperatures $\stackrel{>}{\sim}10$~MK and ratios of X-ray to bolometric luminosity, $L_X/L_{\rm bol}$, at or near coronal saturation levels (i.e., $L_{\rm X}/L_{\rm bol}\sim 10^{-3}$) given the spectral type and bolometric luminosities of the companions. 
This led \citet{2010ApJ...721.1820M} to suggest that the companions were spun-up during the evolution of the CSPN, perhaps by past wind accretion, exchange of orbital energy, or synchronization of the binary orbit and stellar rotation. 
However, in the case of the Helix central star --- which is the brightest and best-studied example of such hard X-ray emission \citep{2001ApJ...553L..55G} --- evidence for a binary companion is only tentative \citep[][]{2001AJ....122..308G}, underscoring the need for caution in invoking companions as sources of the hard X-rays from CSPNe.
\setcounter{footnote}{0}
\footnotetext{Department of Physics \& Astronomy, Vanderbilt University, Nashville, TN}
\footnotetext{Center for Imaging Science and Laboratory for Multiwavelength Astrophysics , Rochester Institute of Technology , Rochester, NY}
\footnotetext{Department of Astronomy, University of Washington , Seattle, WA}
\footnotetext{Department of Physics, Technion, Israel}
\footnotetext{Department of Physics \& Astronomy, University of Rochester, Rochester, NY}
\footnotetext{Observatorio Astronomico Nacional, Alcala de Henares, Spain }
\footnotetext{Academia Sinica, Institute of Astronomy and Astrophysics, Taipei, Taiwan} 
\footnotetext{Instituto de Astrof\'iscia de Canarias, Tenerife, Spain}
\footnotetext{Departamento de Astrof\'isica, University de La Laguna, Tenerife, Spain }
\footnotetext{Department of Physics \& Astronomy and Macquarie Research Centre for Astronomy, Astrophysics, \& Astrophotonics, Macquarie University, Sydney, NSW 2109, Australia}
\footnotetext{Instituto de Astrof\'isica, IAA-CSIC, Granada, Spain}
\footnotetext{Departamento de Fiscia, Universidad de Atacama, Copiap\'o, Chile} 
\footnotetext{Instituto de Astronomia, Universidad Nacional Autonoma de Mexico, Campus Ensenada, Apdo. Postal 22860, Ensenada, B. C., Mexico}
\footnotetext{South African Astronomical Observatory, PO Box 9, Observatory, 7935, South Africa}
\footnotetext{Southern African Large Telescope Foundation, PO Box 9, Observatory, 7935, South Africa}
\footnotetext{NSF Astronomy and Astrophysics Fellow, Center for Computational Relativity and Gravitation, Rochester Institute of Technology, Rochester, NY }
\footnotetext{Australian Astronomical Observatory, PO Box 296, Epping, NSW 2121, Australia}
\footnotetext{Jet Propulsion Laboratory, MS 183-900, California Institute of Technology, Pasadena, CA 91109, USA}
\footnotetext{Leibiniz Institute for Astrophysics Potsdam (AIP), An der Sternwarte 16, D-14482 Potsdam, Germany }
\footnotetext{Columbia Astrophysics Laboratory, Columbia University, New York, NY 10027}
\footnotetext{Department of Physics \& Astronomy, University of Denver, Denver, CO 80208}
\footnotetext{Departmamento de F\'isica Te\'orica, Universidad Aut\'onoma de Madrid, Cantoblanco 28049 Madrid, Spain }
\footnotetext{School of Physics and Astronomy, University of Manchester, Manchester M13 9PL, UK}
\setcounter{footnote}{0}

Alternatively, instabilities in the line-driven stellar winds from CSPNe may also produce 
X-ray emitting shocks in the immediate vicinity of the central star, as has been theorized to explain X-ray emission in O-stars \citep{1997A&A...322..878F}. 
Indeed, \citet{2011MNRAS.417.2440H} have demonstrated the importance of such wind-shock-generated X-ray emission on the photospheric ionization structure in a sample of H-rich CSPNe with far-UV, UV, and optical spectroscopy. 
They find that X-ray fluxes from shocks are necessary to explain the UV emission from high ionization species of oxygen (\ion{O}{6} and \ion{O}{7}) in the coolest CSPNe ($T_{\rm eff}\stackrel{<}{\sim}45$~kK) and can also play a role at higher CSPN temperatures. 
\citet{2013arXiv1307.2948K} have also included X-ray fluxes from shocks in their modeling of central star spectra. 
Their model does not require wind clumping, instead utilizing a self-consistent hydrodynamic method for determining the CSPN and wind parameters. 
In both examples, the authors stress the need for X-ray flux to reproduce the observed UV spectra; hence, the origin of many or perhaps most hard-X-ray emitting CSPNe may be shocks in the circumstellar environment.  
\citet{2013A&A...553A.126G} studied the UV spectra of several CSPNe observed by the Far Ultraviolet Spectroscopic Explorer (FUSE) and found that CSPNe with variable P Cygni profiles can show high ionization potential ions (e.g., \ion{O}{6}) even though the CSPN effective temperatures are too low to to yield such high ionization states.
The authors suggested that Auger ionization from X-ray emission associated with shocks in the CSPN stellar winds are responsible for these highly ionized ions, as is the case in massive OB stars.

Additional explanations for hard X-rays from CSPNe potentially offer novel explorations of other physical processes: 
\begin{itemize}
\item A low-level rate of accretion of PN material, so-called "back flow" \citep{2001MNRAS.328.1081S}, could explain the X-ray emission. 
In the presence of a strong stellar wind spherical accretion is unlikely and for back flow to occur, the material must be dense and limited to a small solid angle  \citep{2001MNRAS.328.1081S}. 
However, for an evolved CSPN like that of the Helix, for which the stellar wind is weak fall back of nebular material cannot be ruled out. 
\item Model atmospheres that account for the effects of departures from local thermodynamic equilibrium (i.e., non-LTE models) predict an excess of ``hard'' X-ray emission for the hottest CSPNe \citep[][also see Appendix]{2003A&A...403..709R}. 
\item A CSPN on its way to becoming a highly magnetic white dwarf \citep{2003ApJ...595.1101S,2005AJ....130..734V} may display magnetic reconnection events and energetic X-ray flaring.
\item If most PNe form from binary systems \citep{2009PASP..121..316D}, then colliding winds between the two stars could provide strong shocks capable of producing X-ray emission, as in symbiotic stars \citep{1997A&A...319..201M} and O star binary systems \citep[e.g.,][]{2011ApJS..194....7N}, but if the secondary is a late type star, the winds may be too weak to support such a scenario.
\end{itemize}
Clearly, for any of these potential explanations for the point-like X-ray emitting sources detected at CSPNe, the binary and wind properties of the CSPN are essential characteristics that can help determine the origin of the X-ray emission.

The Chandra Planetary Nebulae Survey (ChanPlaNS), described in Papers I \& II, has resulted in detections of point-like sources of X-ray emission in the 0.3-3.0 keV band for 20 of 59 PNe observed thus far\footnote{The point-like source of X-ray emission detected from the central star of M 27 (Paper I) is not included among these 20 detected CSPNe, as its photons all have energies $<$0.3 keV.}. 
In this third paper of ChanPlaNS, we explore the X-ray emission from central stars in greater detail. 
We evaluate emission models against the observed X-ray characteristics (\S2.2 \& \S2.3). 
We then use these models to analyze the spectra of the 20 X-ray emitting CSPNe (\S2.4 \& \S2.5). 
The results of our spectral analysis are presented in \S3. 
We have considered these results when estimating upper limits for the 39 undetected CSPNe (\S3.7). 
Finally, we present a summary and discussion the origin of point-like sources of X-ray emission from CSPNe (\S4).

\section{Data Analysis}

Our observations are comprised of archival and targeted observations performed by the Chandra X-ray Observatory (CXO), typically with the back-illuminated chip (S3) of the Advanced CCD for Imaging Spectroscopy (ACIS), which provides optimal soft X-ray sensitivity. 
We follow all recommended data preparation steps and use source detection algorithms to find X-ray source coordinates, count rates, and limits. 
These steps and procedures are described in the ChanPlaNS overview paper \citep{2012AJ....144...58K}. 
In the following, we detail the various analyses employed specifically for the study of point-like X-ray sources detected among the ChanPlaNS PNe.

\subsection{Data Preparation}

We have extracted source and background X-ray count rates, median energies, and spectra for all observed central stars using the CIAO X-ray data analysis software package  \citep[CIAOv4.6;][]{2006SPIE.6270E..60F}. 
Source and background extraction regions were selected to avoid contamination from field X-ray sources and, when present, diffuse X-ray emission. 
Some X-ray point sources are embedded within extended X-ray emission (e.g., NGC 6543, NGC 2392) and we discuss these cases in detail in \S2.5. 
To better isolate the point-like sources we employed the sub-pixel repositioning (SER) algorithm \citep{2003ApJ...590..586L}. 
By employing this algorithm, we decrease the FWHM of the PSF by 40-70\% \citep{2003ApJ...590..586L,2004ApJ...610.1204L,2011ASPC..442..139A}, thereby allowing us to use the smallest source extraction region possible.
Nevertheless,  diffuse X-ray emission may still contaminate the point source spectra, as we further discuss in \S2.5.
The off-axis observation of LoTr 5 requires a larger extraction region; for further details we refer the reader to \citet{2010ApJ...721.1820M}. 

\subsection{Families of Models Considered}

Since the central stars have to be hot enough to ionize their nebulae ($T_{*}>30$~kK), we first attempted to model the X-ray emission with photospheric models. 
The photospheric models we considered included blackbody distributions, solar abundance NLTE models, and NLTE atmosphere models with PG1159-like abundances \citep[see][]{2003A&A...403..709R}.  
However, as demonstrated in Appendix~\ref{appendixA}, in most cases these photospheric mechanisms are unlikely to contribute significantly to X-ray emission in the 0.3-8.0 keV energy band. 
However, photospheric contributions to X-ray emission at energies $\stackrel{<}{\sim}$ 0.3 keV are definitely a possibility. 
Indeed, all the X-ray photons detected from the central star of M 27 have energies $\stackrel{<}{\sim}$ 0.3 keV and are therefore likely due to the $\sim$130 kK central star photosphere. 
However, since the standard calibrations provided by the Chandra X-ray Center are not reliable for energies less than 0.3 keV \citep[see discussion in][]{2013ApJ...766...26M}, we only consider models that can produce photons with energies $\geq0.3$~keV.
In Appendix~\ref{appendixA}, we demonstrate that only the hottest, H-deficient, high-gravity central stars seem capable of producing detectable X-ray flux for photon energies $\geq0.3$ keV. 
Yet, even in these few instances, it is necessary to introduce an additional source of X-ray emission to reproduce the X-ray spectra for energies $\geq0.3$ keV \citep{2007ApJ...670..442H,2013ApJ...766...26M}.

We conclude that a purely photospheric origin is difficult to reconcile with the ChanPlaNS observations of X-rays from CSPNe. 
Hence, for the remainder of our analysis, we explore optically-thin thermal plasma as the most promising emission model for the X-ray emission with photon energies $\geq$0.3 keV detected from CSPNe (as demonstrated in the following section). 
It is important to note that in choosing to focus on thermal plasma models, we have specified the emission model, but not the physical origin of the X-ray emission, since such models can be used to describe accretion, wind shocks, and coronal emission. 

In the analysis that follows, we adopt an optically-thin thermal plasma model \citep[APED/APEC;][]{2012ApJ...756..128F} parameterized by its abundances, temperatures, and emission measure. 
We assume solar abundances  \citep{1989GeCoA..53..197A}  but note that deviations from solar abundances can affect the X-ray spectrum if the X-ray emission originates from highly-ionized emission lines in the metal-rich winds (see \S\ref{discussion}).
We include the effects of intervening absorption --- which may be due to circumstellar and/or interstellar material --- by means of the \verb+wabs+ absorption model \citep{1983ApJ...270..119M}. 

\begin{figure}
\centering
\includegraphics[scale=0.55]{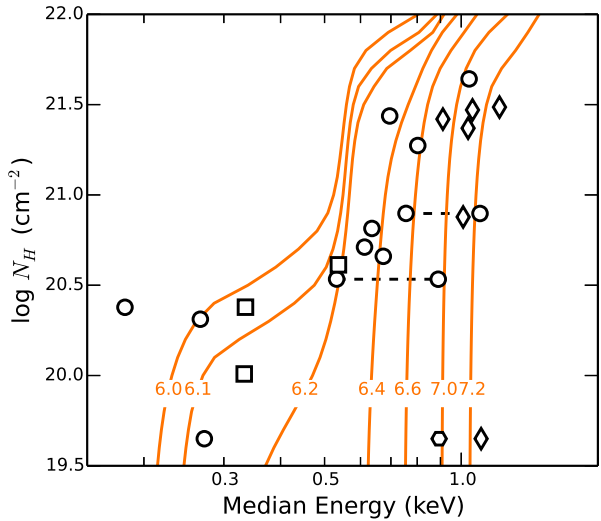}
\caption{Plot of $N_H$ versus median energy for absorbed thermal plasma models and for point-like X-ray emitting sources detected in the ChanPlaNS survey (Papers I \& II). 
Plasma temperatures, log~$T_{\rm X}$ (K), are labeled on the model curves.
Known binary CSPNe and hot CSPNe with WR or PG1159 spectral type have been identified as diamond and square symbols, respectively. 
The CSPN of the Helix Nebula is identified by a hexagon symbol. 
The model-corrected median energies of NGC 2392 and NGC 6543 (see \S\ref{embedsec}) are  connected to the lower-energy, uncorrected points by a dashed line.
The left-most point in this figure is the central star of M 27, which produces no photons above 0.3 keV and is the only source on this diagram characteristic of the ``pure'' X-ray emission of a hot photosphere. 
Objects with little to no extinction (as measured by $c_{H\beta}$) are placed at $\log N_H ({\rm cm}^{-2}) \sim 19.6$.
\label{medenhfig}}
\end{figure}

\subsection{Median Energy as a Proxy\label{medeproxy}}

To verify that the X-ray characteristics of CSPNe are best modeled by an optically-thin thermal plasma, we considered the relevant general parameter space of the absorbed plasma models via a grid of model calculations. 
For ranges of intervening hydrogen column density ($N_H$) from $10^{19}-10^{22} {\rm ~cm}^{-2}$ and characteristic plasma temperature ($T_{\rm X}$) from 1 to 40 MK, with solar abundances, we convolved the plasma models with representative Chandra/ACIS effective area and response matrix calibration files to produce a grid of X-ray spectra. 
We then determined the median energy of each resulting X-ray spectral distribution to find the locus of a given $T_{\rm X}$ in a plot of $N_H$ versus Median Energy.
The resulting loci are plotted in Figure~\ref{medenhfig} and show the behavior of the median energies that would be measured from these absorbed thermal plasma models. 
Positions of the CSPNe are also indicated in Figure~\ref{medenhfig}, with their values of $N_H$ estimated from the $c_{H\beta}$ logarithmic extinction coefficient determined in \citet{2008PhDT.......109F} and \citet{2013MNRAS.431....2F}. 
We used the relationships $E_{B-V} = c_{H\beta} ( 0.61+0.024 c_{H\beta} )$, adopted from \citet{1985PASP...97..700K}, $A_V = R_{V} E_{B-V}$ with $R_{V} = 3.1$, and $N_H = 1.8\times10^{21} {\rm ~mag}^{-1}{\rm ~cm}^{-2} A_V$ \citep{1978ApJ...224..132B} to convert $c_{H\beta}$ to $N_H$. 
For cases where $c_{H\beta} = 0$ we have set $\log N_H ({\rm cm}^{-2}) = 19.6$.
Uncertainties on estimates of the optical extinction ($c_{H\beta}$) range from a few percent up to 50\% \citep[e.g.,][]{1992A&AS...94..399C}, leading to uncertainties of up to 30\% in $E_{B-V}$. 
However, it is likely that the dominant source of uncertainty in the resulting adopted values of $N_H$ is the $A_V$ to $N_H$ conversion factor (i.e., essentially the dust/gas ratio in the absorbing medium). 
Indeed, it is important to point out that the $c_{H\beta}$ measurements are determined from the Balmer decrement of the nebular emission, hence are representative of the extinction between the observer and the hydrogen recombination zone of the nebula. 
Additional absorption of X-ray emission by the ionized material between the central star and this hydrogen recombination zone nebula is possible. 
Such an effect is presumably stronger in the most compact nebulae, for which the nebular densities are highest, and becomes less important for larger nebulae.
Nevertheless, Figure~\ref{medenhfig} indicates that absorbed thermal plasma emission models can reproduce the observed distributions of median energies and $N_H$ that are characteristic of point-source X-ray emission from PNe.
 
\begin{figure*}
\centering
\includegraphics[scale=0.6]{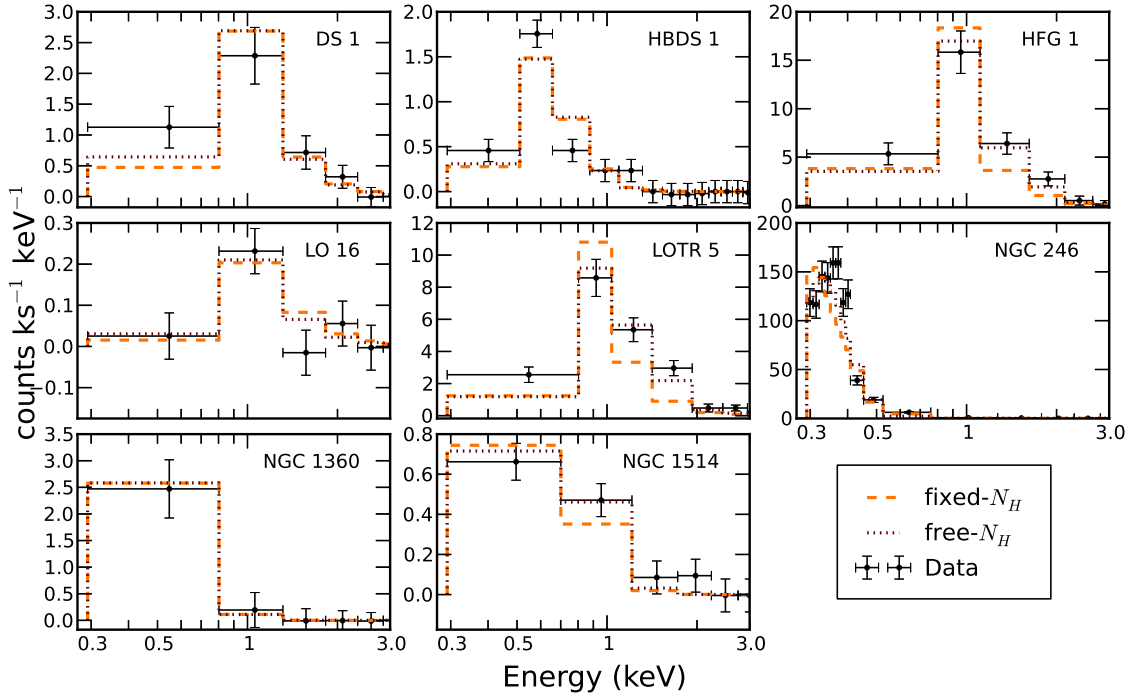}
\caption{Background-subtracted spectra of  CSPNe that are compact X-ray sources. 
We have performed spectral fitting with an absorbed, single temperature, optically-thin thermal plasma, as described in the text (see \S\ref{SecSpecFit}). 
The dashed line indicates the best-fit model with $N_H$ set to the value derived from $c_{H\beta}$ measurements, while the dotted line indicates the best-fit model with $N_H$ left as a free parameter.  
In most cases, the two models agree or are indistinguishable.
\label{spectrafig} }
\end{figure*}

\begin{figure*}
\figurenum{2 ({\it continued})}
\centering
\includegraphics[scale=0.6]{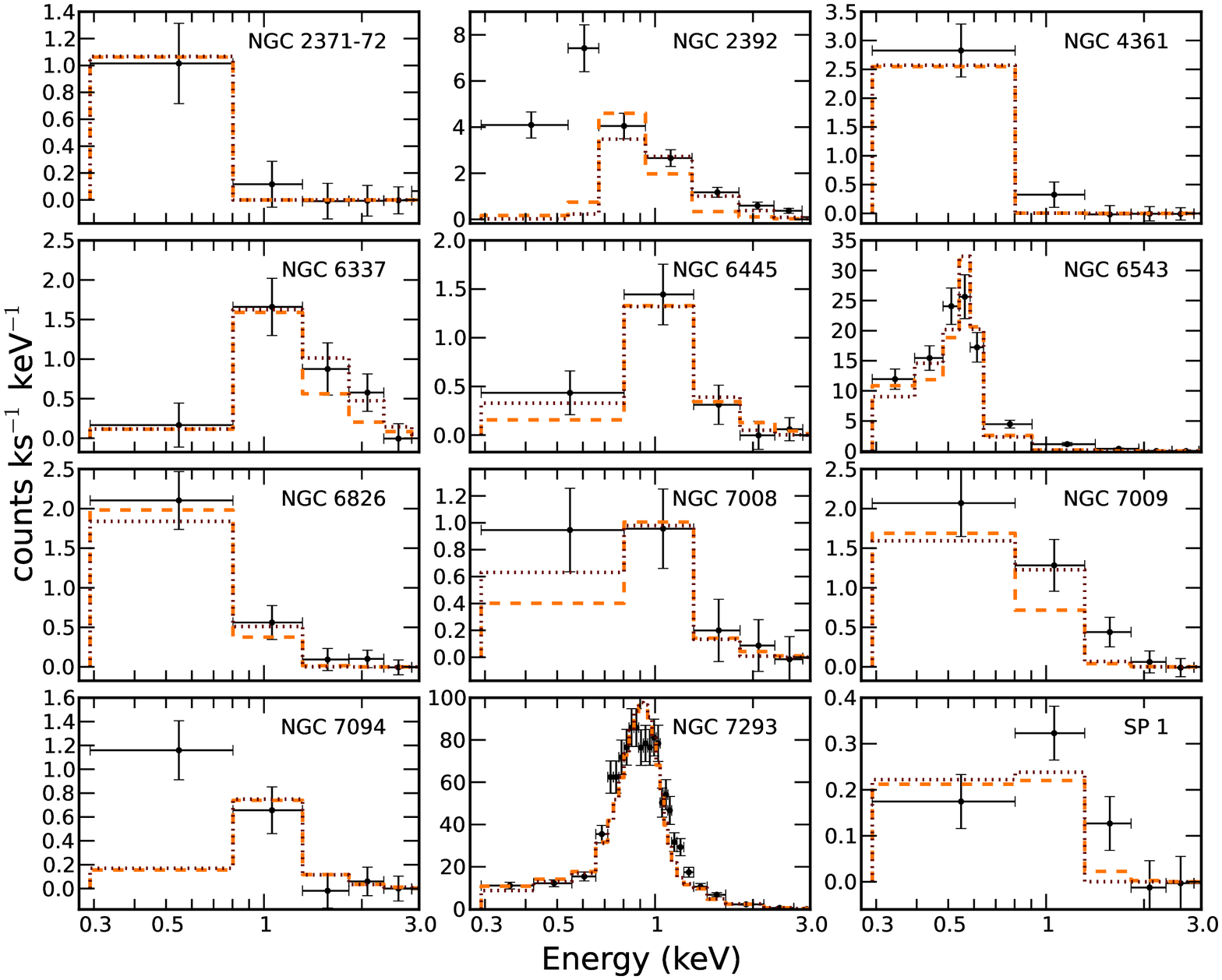}
\caption{As in previous figure. }
\end{figure*}

\subsection{Spectral Fitting\label{SecSpecFit}} 

We analyzed our extracted spectra in XSPEC \citep[version 12.7.1;][]{1996ASPC..101...17A} by fitting  optically-thin isothermal plasma models with solar abundances and suffering intervening absorption (characterized by the column density of H atoms, $N_H$). 
It is important to note that our observations and detections are generally photon-starved. 
The source photons (background-subtracted) number as few as 6, with a median number of detected photons for all sources of 36 (see Table 3 in both Papers I \& II). 
Only 6 sources have more than 100 photons detected. 
Due to the low counts, we find a combination of Churazov weighting and Monte Carlo Markov Chains (MCMC) provides the best balance of spectral fit constraints and ``goodness'' estimation. 
We used 10,000 steps in our MCMC analysis for each spectral fit and studied the $\chi^2$ and model parameter ($N_H$, $T_{\rm X}$, and model normalization) distributions to identify ill-fitting models. 

We performed spectral fits over the 0.3-3.0 keV energy range for two alternative models, one with $N_H$ set to a value calculated from the observed value of $c_{H\beta}$ (referred to as the ``fixed-$N_H$'' model and ``Model 1'') and another with $N_H$ as a free parameter (referred to as the ``free-$N_H$'' model and ``Model 2''). 
The results of our spectral fitting for both models are presented in Figure~\ref{spectrafig} and Table~\ref{specfitstable}. 
Table~\ref{specfitstable} provides the model parameters ($N_H$, $T_X$, and model normalization) and fluxes and luminosities derived by integrating the best-fit model over the 0.3-3.0 keV energy range. 
The observed X-ray flux, $F_{\rm X,obs}$, includes the effects of intervening absorption while the source, or intrinsic, X-ray flux, $F_{\rm X,src}$, does not. 
The X-ray luminosity, $L_{\rm X}$, is calculated from $F_{\rm X,src}$ using $L_{\rm X}=4\pi F_{\rm X,src} D^{2}$ for the distances presented in Table~\ref{cspnebasic}. 
Reported errors in Table~\ref{specfitstable} are 90\% confidence ranges derived from the MCMC distributions for each best-fit parameter with an additional 30\% uncertainty in distance propagated into the error calculation of X-ray luminosity. 

\begin{figure}
\centering
\includegraphics[scale=0.55]{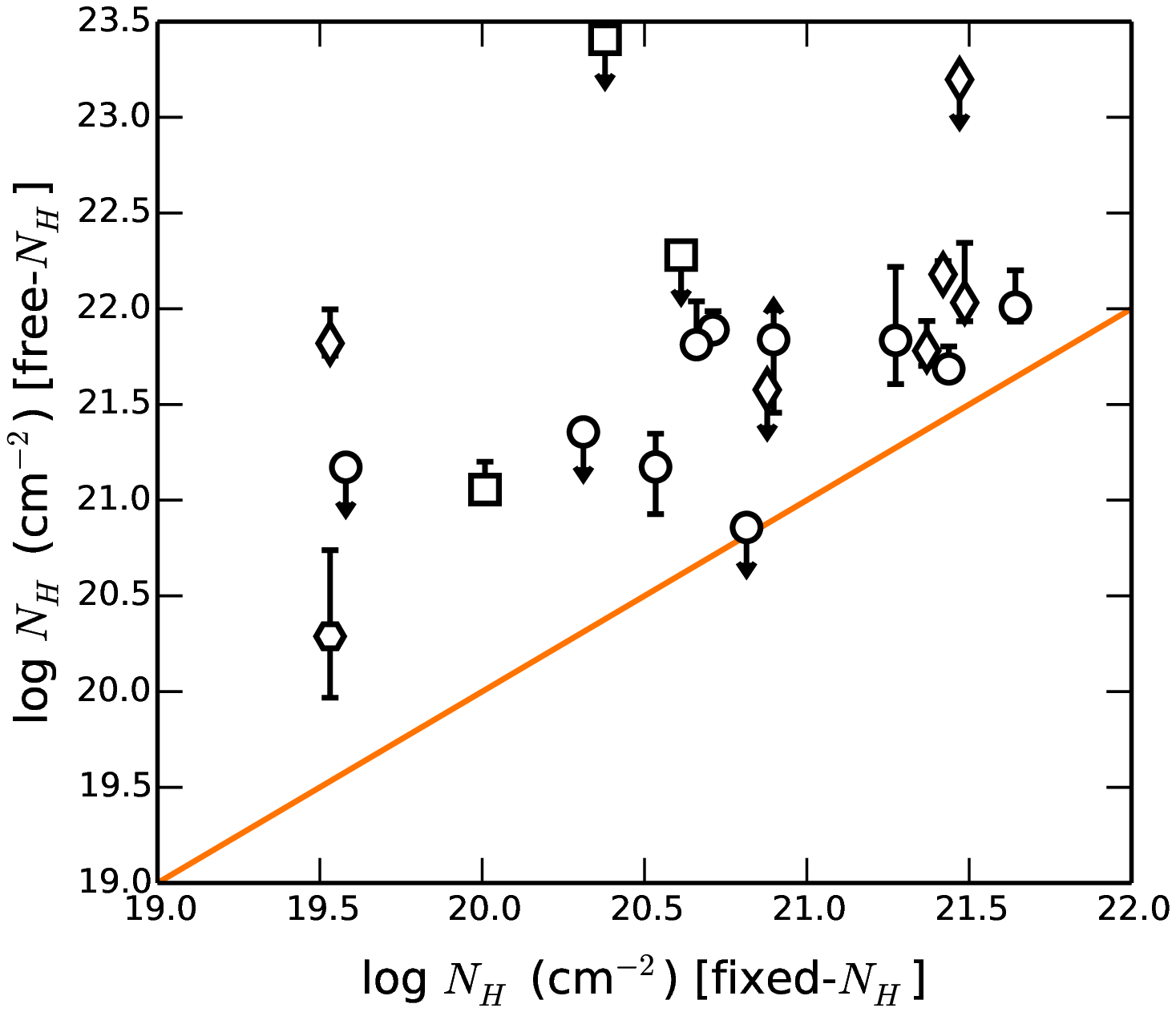}
\caption{Column densities derived from $c_{H\beta}$ measurements (abscissa) and spectral fitting (ordinate).
Known binary CSPNe, hot CSPNe with WR or PG1159 spectral type, and the Helix CSPN have been identified as in Figure~\ref{medenhfig}. 
\label{nhfits}}
\end{figure}

\begin{figure} 
\includegraphics[scale=0.55]{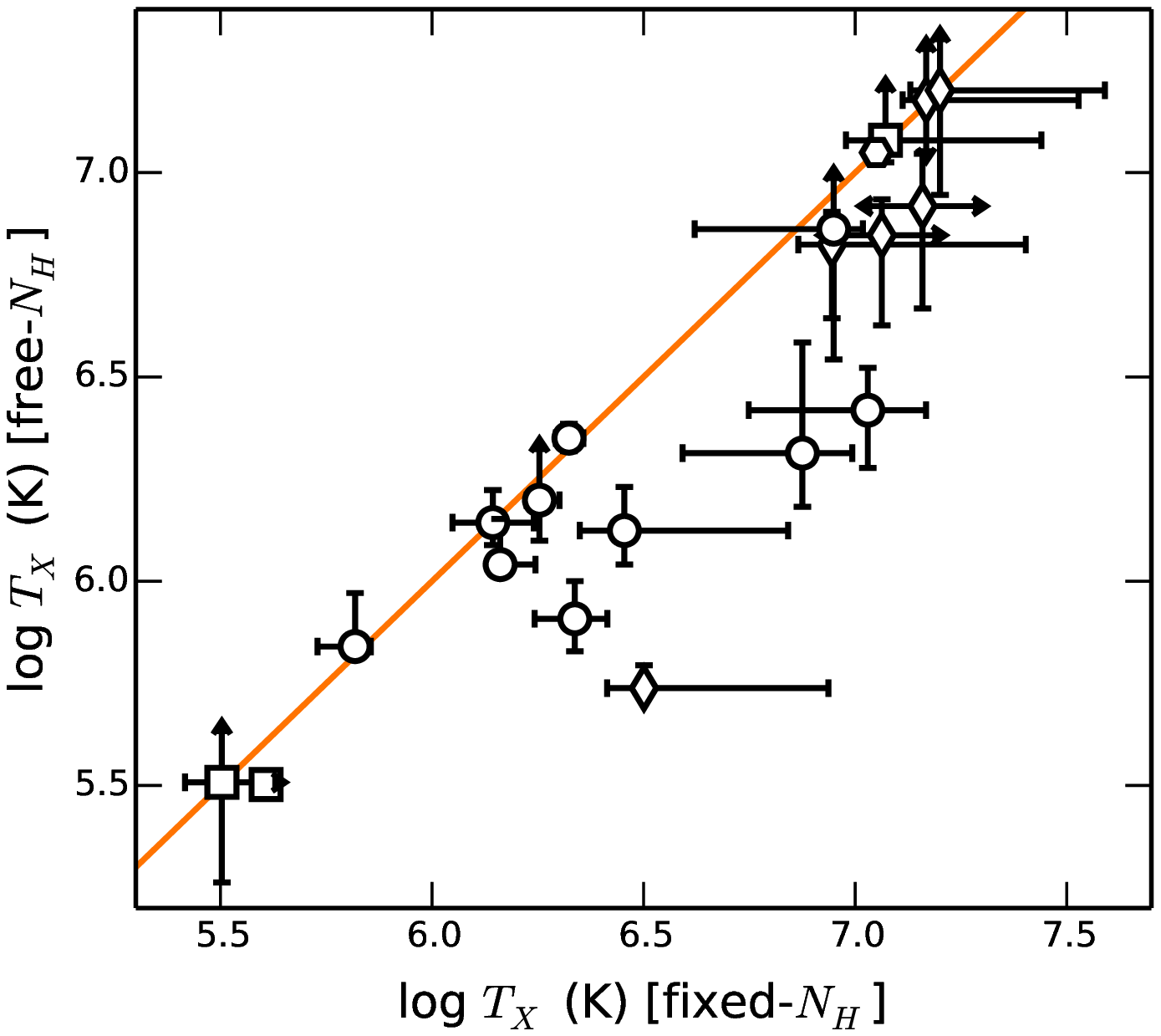}
\includegraphics[scale=0.55]{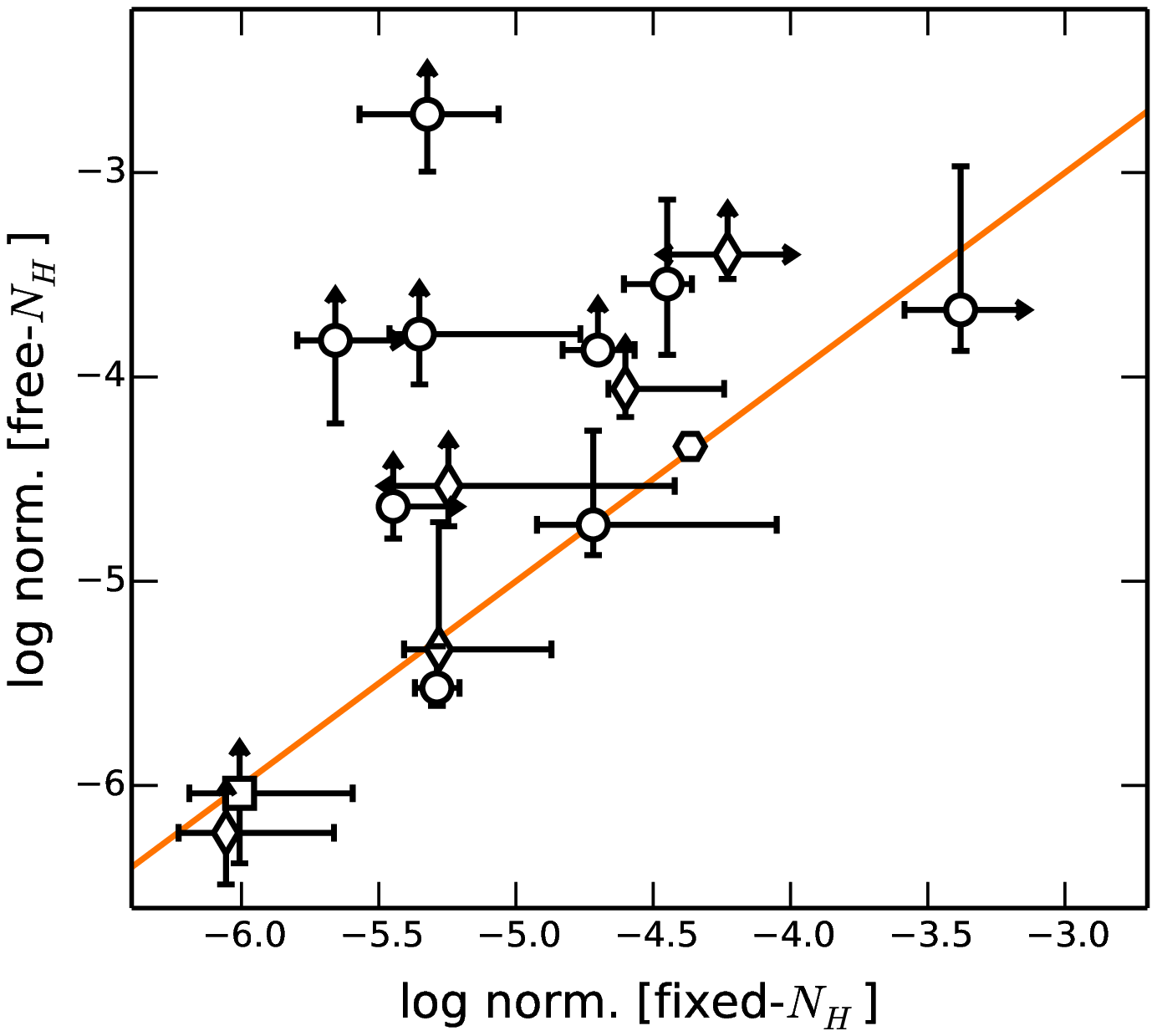}
\caption{Plasma model parameters derived from spectral fits of our two (``free-$N_H$'' and ``fixed-$N_H$''; \S\ref{SecSpecFit}) models. 
Known binary CSPNe, hot CSPNe with WR or PG1159 spectral type, and the Helix CSPN have been identified as in Figure~\ref{medenhfig}. 
The plot range is limited for clarity, causing the data points for NGC 246, NGC 2371-72, NGC 6826, and SP 1 to fall outside of the plot. 
Typically, such a lack of constraint on model normalization occurs when $N_H$ is left as a free parameter (see text and Table~\ref{specfitstable}). 
\label{plasmafits}}
\end{figure}

In Figures~\ref{nhfits} and \ref{plasmafits} we compare the parameters of the fixed-$N_H$ and free-$N_H$ models.
We generally find good agreement.
For those few PNe with ``reliable'' $N_H$ determined from spectral fitting, there is a tendency for $c_{H\beta}$ to under-predict $N_H$, which could indicate either (a) extra absorption and/or (b) the higher metal content of the nebular material (Figure~\ref{nhfits}, also see discussion in \S\ref{medeproxy}). 
Some fits have large uncertainties, arising from our attempt to constrain $N_H$. 
Often the uncertainty in $N_H$ leads to a comparable uncertainty in the model normalization (Figure~\ref{plasmafits}), and hence the model flux and luminosity estimates.
A few free-$N_H$ models predict X-ray luminosities, $L_X$, in the 0.3-3.0 keV range, that are similar to the central star bolometric luminosities, i.e., $L_{\rm X}/L_{\rm bol}\sim1$, which is clearly unphysical.
For certain spectra with very few counts, like that of the CSPN of Lo 16, we are unable to constrain any of the model parameters.
We consider all spectral fits further and on a case by case basis in \S\ref{resultssection}. 

\begin{deluxetable*}{llccccccccccc}
\tablewidth{0pt}
\tabletypesize{\scriptsize}
\tablecaption{X-ray Spectral Fits for Compact Sources in Planetary Nebulae \label{specfitstable}} 
\tablehead{
 \colhead{Object} & \colhead{Mo-} & \colhead{log $N_H$}  & \colhead{$T_{\rm X}$}  & \colhead{log norm} & \colhead{log $F_{X,{\rm obs}}$} & \colhead{log $F_{X,{\rm src}}$} & \colhead{log $L_X$} \\
& \colhead{del} &  \colhead{(cm$^{-2}$)} & \colhead{(MK)} & \colhead{({\rm cm}$^{-3}$)} & \colhead{(erg s$^{-1}$ cm$^{-2}$)} & \colhead{(erg s$^{-1}$ cm$^{-2}$)} & \colhead{(erg s$^{-1}$)} 
}
\startdata
DS 1                     & 1 &  20.88 & 14.72$_{-1.89}^{+12.20}$ & -5.28$_{-0.13}^{+0.41}$ & -14.07$_{-0.11}^{+0.14}$ & -13.97$_{-0.13}^{+0.41}$ & 29.83$_{-0.29}^{+0.49}$ \\ 
 & 2 & $<$21.6 & 15.04$_{-170.60}^{+727.64}$ & -5.33$_{-0.02}^{+0.62}$ & -14.04$_{-0.32}^{+0.14}$ & -14.04$_{-0.02}^{+0.62}$ & 29.77$_{-0.26}^{+0.67}$ \\ 
HBDS 1                   & 1 &  20.81 & 2.11$_{-0.15}^{+0.17}$ & -5.29$_{-0.08}^{+0.08}$ & -14.32$_{-0.06}^{+0.05}$ & -14.06$_{-0.08}^{+0.08}$ & 29.82$_{-0.27}^{+0.27}$ \\ 
 & 2 & $<$20.9 & 2.24$_{-0.16}^{+0.18}$ & -5.52$_{-0.09}^{+0.20}$ & -14.26$_{-0.13}^{+0.13}$ & -14.26$_{-0.09}^{+0.20}$ & 29.62$_{-0.27}^{+0.33}$ \\ 
HFG 1                    & 1 &  21.37 & 8.81$_{-1.61}^{+9.30}$ & -4.60$_{-0.06}^{+0.36}$ & -13.42$_{-0.07}^{+0.09}$ & -13.09$_{-0.06}^{+0.36}$ & 30.54$_{-0.27}^{+0.44}$ \\ 
 & 2 & 21.78$_{-0.08}^{+0.16}$ & 6.67$_{-2.77}^{+1.23}$ & -4.06$_{-0.14}^{+0.97}$ & -13.36$_{-0.08}^{+0.05}$ & -12.56$_{-0.14}^{+0.97}$ & 31.07$_{-0.29}^{+1.00}$ \\ 
LO 16                    & 1 &  21.47 & 15.87$_{-2.56}^{+14.24}$ & -6.06$_{-0.17}^{+0.39}$ & -15.08$_{-0.16}^{+0.21}$ & -14.80$_{-0.17}^{+0.39}$ & 29.13$_{-0.31}^{+0.47}$ \\ 
 & 2 & $<$23.2 & 15.90$_{-9.37}^{+725.12}$ & -6.23$_{\rm n/a}^{\rm n/a}$ & -15.09$_{\rm n/a}^{\rm n/a}$  & -14.98$_{\rm n/a}^{\rm n/a}$  & 28.95$_{\rm n/a}^{\rm n/a}$  \\ 
LOTR 5$^{\dagger}$                   & 1 &  19.53 & 11.56$_{-12.22}^{+25.85}$ & -4.23$_{-0.72}^{+1.18}$ & -12.79$_{-0.10}^{+0.26}$ & -12.79$_{-0.72}^{+1.18}$ & 30.69$_{-0.77}^{+1.21}$ \\ 
 & 2 & 21.82$_{-0.06}^{+0.18}$ & 7.03$_{-3.57}^{+1.42}$ & -3.40$_{-0.12}^{+1.72}$ & -12.73$_{-0.08}^{+0.04}$ & -11.90$_{-0.12}^{+1.72}$ & 31.58$_{-0.29}^{+1.74}$ \\ 
NGC 246$^{\dagger\dagger}$                  & 1 &  20.01 & 0.41$_{-0.00}^{+0.00}$ & -0.59$_{-0.08}^{+0.08}$ & -12.28$_{-0.03}^{+0.03}$ & -12.16$_{-0.08}^{+0.08}$ & 31.31$_{-0.27}^{+0.27}$ \\ 
 & 2 & 21.06$_{-0.06}^{+0.15}$ & 0.32$_{-0.01}^{+0.00}$ & 1.44$_{-0.16}^{+0.87}$ & -12.45$_{-0.07}^{+0.43}$ & -11.13$_{-0.16}^{+0.87}$ & 32.35$_{-0.30}^{+0.91}$ \\ 
NGC 1360                 & 1 &  19.53 & 1.39$_{-0.31}^{+0.31}$ & -4.72$_{-0.20}^{+0.67}$ & -13.84$_{-0.12}^{+0.19}$ & -13.82$_{-0.20}^{+0.67}$ & 29.41$_{-0.33}^{+0.72}$ \\ 
 & 2 & $<$21.2 & 1.39$_{-0.18}^{+0.26}$ & -4.72$_{-0.15}^{+0.46}$ & -13.84$_{-0.25}^{+0.18}$ & -13.83$_{-0.15}^{+0.46}$ & 29.41$_{-0.30}^{+0.53}$ \\ 
NGC 1514                 & 1 &  21.44 & 1.79$_{-0.12}^{+0.20}$ & -4.70$_{-0.13}^{+0.13}$ & -14.55$_{-0.06}^{+0.05}$ & -13.58$_{-0.13}^{+0.13}$ & 29.64$_{-0.29}^{+0.29}$ \\ 
 & 2 & 21.69$_{-0.03}^{+0.12}$ & 1.58$_{-0.36}^{+0.00}$ & -3.87$_{-0.05}^{+1.80}$ & -14.54$_{-0.06}^{+0.03}$ & -12.85$_{-0.05}^{+1.80}$ & 30.36$_{-0.26}^{+1.82}$ \\ 
NGC 2371-72$^{\dagger\dagger}$              & 1 &  20.38 & 0.32$_{-0.06}^{+0.00}$ & -0.07$_{-0.04}^{+3.10}$ & -12.91$_{-0.21}^{+0.19}$ & -12.62$_{-0.04}^{+3.10}$ & 31.76$_{-0.26}^{+3.11}$ \\ 
 & 2 & $<$23.4 & 0.32$_{-0.18}^{+1.02}$ & -0.39$_{\rm n/a}^{\rm n/a}$  & -12.84$_{\rm n/a}^{\rm n/a}$  & -12.84$_{\rm n/a}^{\rm n/a}$  & 31.54$_{\rm n/a}^{\rm n/a}$  \\ 
NGC 2392$^{\dagger}$                 & 1 &  20.90 & 8.90$_{-6.73}^{+1.42}$ & -5.45$_{-0.03}^{+1.68}$ & -14.06$_{-0.04}^{+0.37}$ & -13.94$_{-0.03}^{+1.68}$ & 30.35$_{-0.26}^{+1.70}$ \\ 
 & 2 & 21.84$_{-0.38}^{+0.20}$ & 7.27$_{-5.34}^{+4.57}$ & -4.63$_{-0.16}^{+1.20}$ & -13.98$_{-0.01}^{+0.14}$ & -13.13$_{-0.16}^{+1.20}$ & 31.16$_{-0.30}^{+1.22}$ \\ 
NGC 4361                 & 1 &  20.31 & 0.66$_{-0.13}^{+0.06}$ & -3.38$_{-0.20}^{+1.83}$ & -13.78$_{-0.10}^{+0.27}$ & -13.58$_{-0.20}^{+1.83}$ & 30.45$_{-0.33}^{+1.85}$ \\ 
 & 2 & $<$21.4 & 0.69$_{-0.04}^{+0.21}$ & -3.67$_{-0.20}^{+0.70}$ & -13.75$_{-0.34}^{+0.37}$ & -13.75$_{-0.20}^{+0.70}$ & 30.28$_{-0.33}^{+0.75}$ \\ 
NGC 6337                 & 1 &  21.49 & 14.43$_{-2.05}^{+57.25}$ & -5.25$_{-0.05}^{+0.82}$ & -14.23$_{-0.06}^{+0.33}$ & -13.92$_{-0.05}^{+0.82}$ & 30.02$_{-0.27}^{+0.86}$ \\ 
 & 2 & 22.03$_{-0.10}^{+0.31}$ & 8.28$_{-4.78}^{+2.45}$ & -4.53$_{-0.20}^{+2.65}$ & -14.09$_{-0.17}^{+0.09}$ & -13.02$_{-0.20}^{+2.65}$ & 30.93$_{-0.33}^{+2.67}$ \\ 
NGC 6445                 & 1 &  21.64 & 10.71$_{-6.95}^{+3.39}$ & -5.35$_{-0.11}^{+0.59}$ & -14.37$_{-0.16}^{+0.12}$ & -13.88$_{-0.11}^{+0.59}$ & 30.48$_{-0.28}^{+0.64}$ \\ 
 & 2 & 22.01$_{-0.08}^{+0.19}$ & 2.62$_{-0.85}^{+0.63}$ & -3.79$_{-0.25}^{+2.49}$ & -14.37$_{-0.18}^{+0.12}$ & -12.47$_{-0.25}^{+2.49}$ & 31.89$_{-0.36}^{+2.50}$ \\ 
NGC 6543$^{\dagger}$                 & 1 &  20.53 & 1.45$_{-0.09}^{+0.28}$ & -4.45$_{-0.16}^{+0.09}$ & -13.69$_{-0.05}^{+0.04}$ & -13.51$_{-0.16}^{+0.09}$ & 30.92$_{-0.30}^{+0.28}$ \\ 
 & 2 & 21.17$_{-0.25}^{+0.17}$ & 1.10$_{-0.06}^{+0.28}$ & -3.55$_{-0.35}^{+0.41}$ & -13.70$_{-0.05}^{+0.12}$ & -12.91$_{-0.35}^{+0.41}$ & 31.52$_{-0.43}^{+0.49}$ \\ 
NGC 6826                 & 1 &  20.71 & 2.17$_{-0.48}^{+0.39}$ & -5.23$_{-0.15}^{+0.28}$ & -14.19$_{-0.13}^{+0.11}$ & -13.99$_{-0.15}^{+0.28}$ & 30.32$_{-0.30}^{+0.38}$ \\ 
 & 2 & 21.89$_{-0.06}^{+0.10}$ & 0.81$_{-0.15}^{+0.17}$ & -1.06$_{-0.36}^{+2.58}$ & -14.25$_{-0.14}^{+0.03}$ & -10.87$_{-0.36}^{+2.58}$ & 33.43$_{-0.44}^{+2.59}$ \\ 
NGC 7008$^{\dagger\dagger}$                 & 1 &  21.27 & 7.50$_{-4.90}^{+2.04}$ & -5.66$_{-0.14}^{+0.99}$ & -14.44$_{-0.15}^{+0.17}$ & -14.15$_{-0.14}^{+0.99}$ & 29.62$_{-0.30}^{+1.02}$ \\ 
 & 2 & 21.84$_{-0.23}^{+0.38}$ & 2.06$_{-0.62}^{+1.28}$ & -3.82$_{-0.41}^{+3.01}$ & -14.38$_{-0.20}^{+0.11}$ & -12.60$_{-0.41}^{+3.01}$ & 31.16$_{-0.48}^{+3.02}$ \\ 
NGC 7009                 & 1 &  20.61 & 2.85$_{-0.69}^{+2.54}$ & -5.32$_{-0.25}^{+0.26}$ & -14.13$_{-0.18}^{+0.12}$ & -13.99$_{-0.25}^{+0.26}$ & 30.41$_{-0.36}^{+0.37}$ \\ 
 & 2 & 21.81$_{-0.06}^{+0.23}$ & 1.33$_{-0.25}^{+0.33}$ & -2.71$_{-0.28}^{+1.33}$ & -14.11$_{-0.19}^{+0.07}$ & -11.87$_{-0.28}^{+1.33}$ & 32.53$_{-0.38}^{+1.36}$ \\ 
NGC 7094$^{\dagger\dagger}$                 & 1 &  20.61 & 11.79$_{-2.54}^{+10.01}$ & -6.01$_{-0.18}^{+0.41}$ & -14.64$_{-0.19}^{+0.16}$ & -14.58$_{-0.18}^{+0.41}$ & 29.79$_{-0.32}^{+0.49}$ \\ 
 & 2 & $<$22.3 & 12.00$_{-1.49}^{+720.12}$ & -6.04$_{-0.34}^{+1.97}$ & -14.62$_{-0.41}^{+0.09}$ & -14.62$_{-0.34}^{+1.97}$ & 29.75$_{-0.43}^{+1.99}$ \\ 
NGC 7293                 & 1 &  19.53 & 11.22$_{-0.28}^{+0.26}$ & -4.37$_{-0.02}^{+0.02}$ & -12.92$_{-0.02}^{+0.02}$ & -12.91$_{-0.02}^{+0.02}$ & 29.85$_{-0.26}^{+0.26}$ \\ 
 & 2 & 20.29$_{-0.32}^{+0.45}$ & 11.20$_{-0.30}^{+0.26}$ & -4.34$_{-0.03}^{+0.04}$ & -12.92$_{-0.02}^{+0.01}$ & -12.89$_{-0.03}^{+0.04}$ & 29.88$_{-0.26}^{+0.26}$ \\ 
SP 1                     & 1 &  21.42 & 3.17$_{-0.64}^{+3.18}$ & -5.63$_{-0.26}^{+0.25}$ & -14.91$_{-0.11}^{+0.10}$ & -14.28$_{-0.26}^{+0.25}$ & 29.90$_{-0.37}^{+0.36}$ \\ 
 & 2 & 22.18$_{-0.03}^{+0.07}$ & 0.55$_{-0.03}^{+0.07}$ & 1.67$_{-0.29}^{+0.71}$ & -14.89$_{-0.14}^{+0.06}$ & -8.94$_{-0.29}^{+0.71}$ & 35.24$_{-0.39}^{+0.75}$  
\enddata
\tablecomments{ Model 1 and Model 2 correspond to the fixed-$N_H$ and free-$N_H$ models, respectively, as discussed in the text. }
\tablenotetext{$\dagger$}{Indicates that we adopt a separate value for the spectral fit. 
In the case of LoTr 5, we adopt the fit from \citet{2010ApJ...721.1820M}, while NGC 2392 and NGC 6543 are found to be contaminated by diffuse X-ray emission and their spectral fits are revisited in \S\ref{embedsec} and reported in Table~\ref{embeddedsources}.}
\tablenotetext{$\dagger\dagger$}{Indicates spectral fits are unreliable for NGC 246, NGC 2371-72, NGC 7008, and NGC 7094, as discussed in \S\ref{resultssection}.} 
\end{deluxetable*}

\subsection{Mitigating contamination of CSPN X-ray spectra by diffuse X-ray emission\label{embedded_section}}

Five ChanPlaNS PNe (NGC 2371-72, NGC 2392, NGC 6543, NGC 6826, and NGC 7009) contain both point-like and diffuse sources of X-ray emission. 
The extracted spectra of the point-like component may be contaminated by the diffuse X-ray emission. 
Here, we describe the use of spatial and spectral properties of the two emission sources to distinguish between them and thereby determine the degree of contamination of the point-like source by the diffuse source. 

\begin{figure}
\centering 
\includegraphics[scale=0.65]{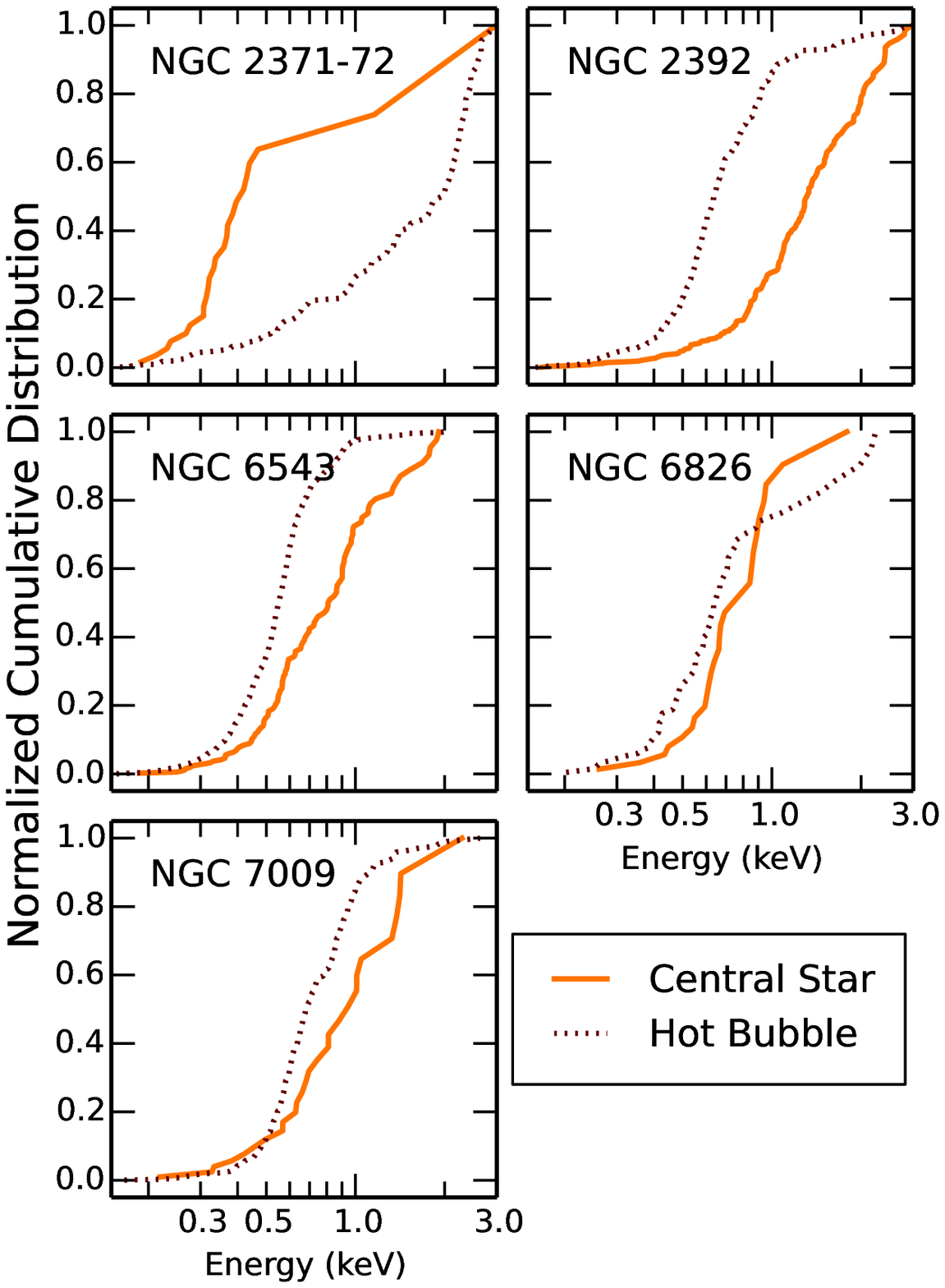}
\caption{
Cumulative distributions of the X-ray emission from the central star (solid line) and  diffuse emission with the central star excised (dotted line).
\label{spectralsepfig1}}
\end{figure}

First, we compare the cumulative distributions of photon energies for the point-like source and for the diffuse source with the point-like source excised (Figure~\ref{spectralsepfig1}). 
These cumulative distributions reveal the stark difference between the central point-like sources in NGC 2371-72, which is very soft compared to the diffuse emission, and in NGC 2392, whose point-like source is very hard compared to the diffuse emission.
Other point-like sources are less distinct from their diffuse counterparts.
To quantify the difference of the cumulative distributions, we perform a two-sided, two-sample K-S test to compare the point-like and diffuse energy distributions (Table~\ref{kstest}). 
The p-values of our K-S tests suggest that all of our point-like sources are distinct with respect to the diffuse X-ray emission, with the point-like sources in NGC 6826 and NGC 7009 showing the most similarity to their diffuse counterparts.

\begin{figure*}
\centering
\includegraphics[scale=0.6]{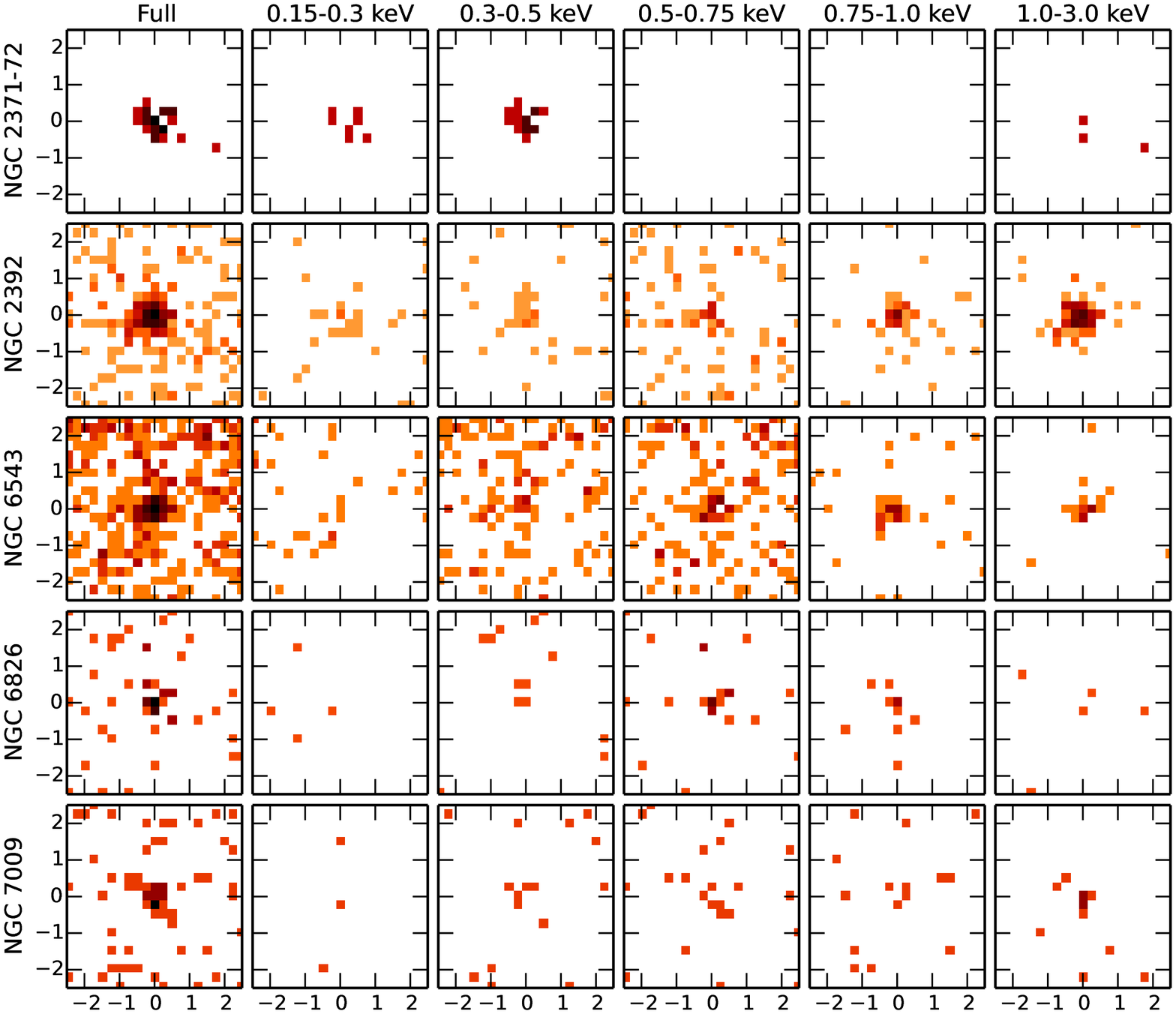}
\caption{
Energy-filtered images centered on the central regions of PNe with CSPN and diffuse X-ray emission. 
These images, which increase in energy from left to right, show where the diffuse and compact sources emit. 
\label{spectralsepfig2}}
\end{figure*}

Next we generated narrow energy-filtered images across the 0.15 keV to 3.0 keV energy range, that are restricted to a small region ($5^{\prime\prime}\times5^{\prime\prime}$) around the central point-like source. 
For a view of the entire nebula and diffuse X-ray emission, we refer the reader to Figure~3 of Papers I \& 2. 
By zooming into the central region, a given sequence of images allows us to study the spatial distribution of the detected photons in each energy range (see Figure~\ref{spectralsepfig2}). 
From these images we conclude that the point-like sources in NGC 2392 and NGC 6543 are highly contaminated by the diffuse component at soft energies $\stackrel{<}{\sim} 0.75$~keV.
In the lower energy bands, NGC 2392 and NGC 6543 appear contaminated by the diffuse emission, but the point-like source peaks at higher energies. 
The point-like source in NGC 2371-72 is not contaminated since its diffuse emission is quite far from the central source.
The low count rate emission detected from the central regions of NGC 6826 and NGC 7009 makes it difficult to determine the degree of contamination, if any. 
Thus, for NGC 6826 and NGC 7009, we cannot make any further quantitative estimates for the degrees of diffuse contamination of their central point-like sources.

\begin{figure*}
\centering
\includegraphics[scale=0.6]{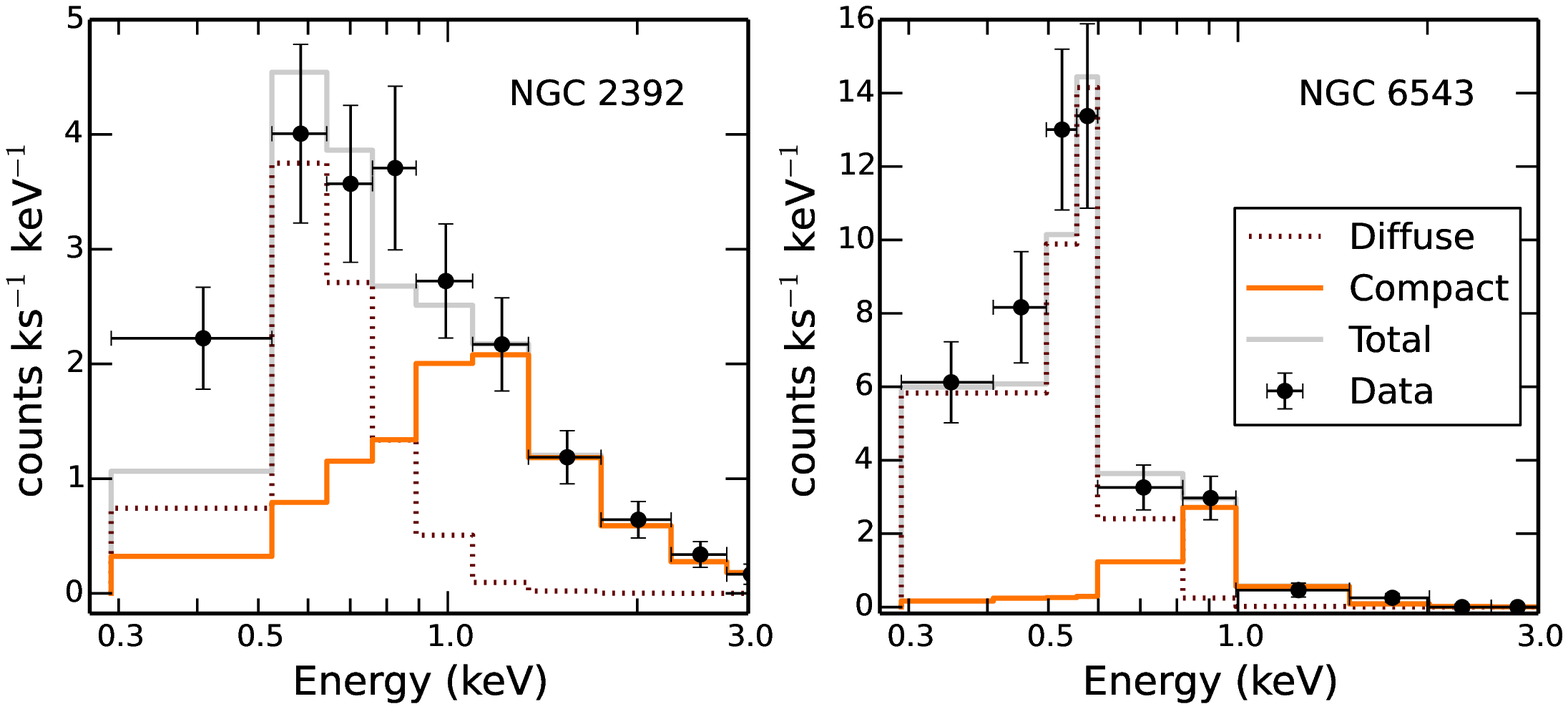}
\caption{
Spectra of the CSPN emission from NGC 2392 (left) and NGC 6543 (right), which are contaminated by diffuse X-ray emission. 
The adopted spectral fit for two plasma temperature model is overlaid along with the individual plasma components as described by the legend. 
\label{spectralsepfig3} }
\end{figure*}

From the narrow-band spectral images (Figure~\ref{spectralsepfig2}) we identified that the point-like sources in NGC 2392 and NGC 6543 could be separated spectroscopically.  
For each point-like source spectrum, we ignored energies $<0.75$~keV, which removes most of the contaminating diffuse X-ray emission, as suggested by Figure~\ref{spectralsepfig2}. 
Next we fit the X-ray spectrum for energies $>0.75$~keV with both the fixed-$N_H$ and free-$N_H$ isothermal plasma models.
Finally, we add the previously fit spectral model from Table~\ref{specfitstable} and refit the entire 0.3 to 3.0 keV X-ray spectrum with this hybrid two-temperature component model. 
The resulting fits are presented in Table~\ref{embeddedsources} and Figure~\ref{spectralsepfig3}. 
The energy distributions of these newly added hard components better reproduce the point-like emission seen in the narrow-band spectral images in Figure~\ref{spectralsepfig2}, leading us to adopt the hard component as representative of the central source, while the soft component represents the contamination by the diffuse X-ray emission.  

\begin{deluxetable*}{lccccccccc}
\tablewidth{0pt}
\tabletypesize{\scriptsize}
\tablecaption{Spectral Separation of Compact and Diffuse Sources \label{kstest}} 
\tablehead{
 & \multicolumn{2}{c}{Central Star} & \multicolumn{2}{c}{Hot Bubble} & \multicolumn{2}{c}{K-S Two-Sample Test}  \\
\colhead{Object} & \colhead{N$_{\rm photons}$} & \colhead{E$_{\rm median}$}  &  \colhead{N$_{\rm photons}$} & \colhead{E$_{\rm median}$} &  \colhead{$\Delta D$} & \colhead{ p-value} }
\startdata
NGC 2371-72 & 24 & 0.33 & 123 & 0.69 & 0.62 & 1.79e-07  \\
NGC 2392 & 195 & 1.05 & 640 & 0.57 & 0.52 & 2.84e-36 \\
NGC 6543 & 115 & 0.59 & 1511 & 0.51 & 0.31 & 2.07e-09  \\
NGC 6826 & 26 & 0.66 & 71 & 0.55 & 0.32 & 2.93e-02 \\
NGC 7009 & 27 & 0.70 & 425 & 0.61 & 0.22 & 1.37e-01 \\
\enddata
\tablecomments{
We provide the properties of the compact and diffuse sources in five PNe that show both types of emission. 
The diffuse characteristics were determined from the total diffuse emission with the central sources excised. 
The two-sample K-S test produces two metrics, the K-S statistic, $\Delta D$, which is the maximum distance between the two normalized cumulative distributions (see Figure~\ref{spectralsepfig1}), and the p-value derived from $\Delta D$. 
Confidence in the rejection of the null hypothesis that both samples are from identical distributions decreases with higher p-values. 
}
\end{deluxetable*}

\begin{deluxetable*}{llccccccccccc}
\tablewidth{0pt}
\tabletypesize{\scriptsize}
\tablecaption{X-ray Spectral Fits for Embedded Compact Sources in Planetary Nebulae \label{embeddedsources}} 
\tablehead{
 \colhead{Object} & \colhead{Component} & \colhead{log $N_H$}  & \colhead{$T_{\rm X}$}  & \colhead{log norm} & \colhead{log $F_{\rm X,{\rm obs}}$} & \colhead{log $F_{\rm X,{\rm src}}$} & \colhead{log $L_{\rm X}$} \\
 & &  \colhead{(cm$^{-2}$)} & \colhead{(MK)} & \colhead{({\rm cm}$^{-3}$)} & \colhead{(erg s$^{-1}$ cm$^{-2}$)} & \colhead{(erg s$^{-1}$ cm$^{-2}$)} & \colhead{(erg s$^{-1}$)} 
}
\startdata
NGC 2392    & Compact &  20.90 & 36.28$_{-7.81}^{+19.64}$ & -4.93$_{-0.08}^{+0.08}$ & -13.78$_{-0.07}^{+0.07}$ & -13.71$_{-0.08}^{+0.08}$ & 30.58$_{-0.27}^{+0.27}$ \\ 
NGC 2392    & Diffuse &  20.90 & 2.18$_{-0.49}^{+0.52}$ & -5.13$_{-0.16}^{+0.30}$ & -14.19$_{-0.07}^{+0.07}$ & -13.89$_{-0.16}^{+0.30}$ & 30.40$_{-0.31}^{+0.40}$ \\ 
NGC 6543    & Compact &  20.53 & 10.07$_{-1.77}^{+2.05}$ & -5.94$_{-0.11}^{+0.15}$ & -14.50$_{-0.06}^{+0.06}$ & -14.44$_{-0.11}^{+0.15}$ & 29.98$_{-0.28}^{+0.30}$ \\ 
NGC 6543    & Diffuse &  20.53 & 1.29$_{-0.16}^{+0.18}$ & -4.63$_{-0.15}^{+0.23}$ & -14.00$_{-0.06}^{+0.06}$ & -13.81$_{-0.15}^{+0.23}$ & 30.61$_{-0.30}^{+0.35}$ \\ 
\enddata
\tablecomments{These nebula feature both compact X-ray sources at the location of the central star and diffuse X-ray emission which we have spectrally separated. Following the discussion in \S\ref{embedded_section}, we present the adopted fixed-$N_H$ models. Diffuse spectral fits are only for the emission contained within the circular aperture used for extracting the central point source and do not represent the total diffuse emission. }
\end{deluxetable*}

\section{Results}\label{resultssection}

To facilitate discussion, we adopt one of the two isothermal  plasma models (fixed-/free-$N_H$) based on the quality of the spectral fit and physical parameters of each model for each CSPN.
In many cases, it is straightforward to choose the fixed-$N_H$ model over free-$N_H$ model because of the lack of constraint on $N_H$ and/or the prediction of an unphysical value of $L_X$ (Table~\ref{specfitstable}). 
Below we further detail the choice of a particular model and begin to evaluate the potential origin(s) for the point-like X-ray emission on a case by case basis, including (where possible) supplementary information on the CSPN (temperature, luminosity, spectral type, winds, binarity, and nebular morphology). 
In Table~\ref{cspnebasic} columns 2-5 provide the central star properties ($D$, $T_{\rm star}$, $L_{\rm bol}$, and spectral type [S.T.]) taken from \citet{2008PhDT.......109F} and Papers~I~\&~II, except as indicated. 
Column 6 gives binary information compiled from \citet{2009PASP..121..316D}, \citet{2008PhDT.......109F}, and updates maintained by D. Jones\footnote{A list of known binary CSPNe is maintained by Dr. David Jones at \url{http://www.drdjones.net/?q=bCSPN}}, for detailed references see Papers I \& II. 
Column 7 gathers information from FUSE \citep{2013A&A...553A.126G} and IUE \citep{1985ApJ...291..237C,1991A&AS...91..325P} UV spectroscopy. 
Columns 8-11 detail the adopted model parameters (Model, $N_H$, $T_{\rm X}$, and $L_{\rm X}$) from Table~\ref{specfitstable} and Table~\ref{embeddedsources}. 
Finally, in the following, we have grouped the CSPNe into a few loosely-defined categories.

\subsection{CSPNe with active or close companions}

\noindent DS 1, HFG 1, Lo 16, LoTr 5, NGC 6337, and Sp 1:
Each of these CSPN is a known binary system with a close binary nucleus (orbital period $\stackrel{<}{\sim}20$~days) or a known rapidly rotating companion (in the case of LoTr 5). 
LoTr 5, DS 1, and HFG 1 were the subject of a detailed-study by \citet{2010ApJ...721.1820M}. 
For DS 1, HFG 1, Lo 16, and Sp 1 we adopt the fixed-$N_H$ model, which better reproduces the spectral energy distributions; furthermore, for these harder X-ray sources, it is more difficult to constrain $N_H$, which is more sensitive to soft X-ray photons. 
For NGC 6337 we adopt the free-$N_H$ model. 
For LoTr 5, neither isothermal plasma model could reproduce the spectrum, so we adopt the two-temperature plasma model from \citet{2010ApJ...721.1820M}. 
There is no evidence for a stellar wind from DS 1 or HFG 1, while LoTr 5 has a weak wind. 
There are no known UV observations of the CSPNe of Lo 16 and NGC 6337. 
We later (\S4.2) argue that magnetically active coronae at the companions are the likely sources of the X-ray emission in all five of these cases.

\subsection{Hot PG1159-type CSPNe} 

\noindent NGC 2371-72, NGC 246, NGC 7094: 
Each of these hot PG1159 central stars have no compelling evidence for close binary companions.
However, all display strong P Cygni wind profiles. 
We cannot constrain $N_H$ for any of these sources and neither the fixed-$N_H$ nor free-$N_H$ models can reproduce the spectra of  NGC 7094 nor NGC 246. 
We interpret this as an indication that these models are poor representations of the X-ray emission. 
The spectra are soft, which might indicate the presence of highly ionized carbon line emission as in the hot PG1159 central stars within K 1-16 \citep{2013ApJ...766...26M} and NGC 246  \citep{2007ApJ...670..442H}. 
That NGC 246 and K 1-16 are both hydrogen deficient PG1159-type central stars led \citet{2013ApJ...766...26M} to suggest that such carbon-rich X-ray emitting plasma is a common feature of this class of CSPNe, which are believed to have undergone a carbon-enhancing third dredge-up \citep{2006PASP..118..183W}. 
Among our sample, NGC 2371-72 is a [WCE]-type central star and part of the \ion{O}{6} sequence \citep{1984ApJ...278..195K} that is believed to be a precursor to the PG1159-type stars \citep{2006PASP..118..183W}. 
The CSPN of NGC 7094 is a member of the rare ``hybrid'' PG1159 central star class \citep{1995A&A...301..545N,2000ApJ...542..957F}; in addition to PG1159-type characteristics, the CSPNe in this class show strong Balmer hydrogen lines \citep{1991A&A...249L..16N}. 
Like NGC 246, the CSPNe of NGC 2371-72 and NGC 7094 feature soft X-ray emission that could be due to carbon-enhanced circumstellar plasmas and, hence, require a detailed treatment that is beyond the scope of this paper. 
Furthermore, the low X-ray plasma temperatures might suggest that NLTE photospheric emission contributes to the emission from these hot CSPNe, as in the case of K 1-16 \citep{2013ApJ...766...26M}. 
We advise caution in interpreting the model-derived parameters for these three sources. 

\subsection{Point-like sources embedded in diffuse emission\label{embedsec}} 

\noindent NGC 2392, NGC 6543, NGC 7009, NGC 6826:
Each of these PN display resolved composite X-ray sources comprised of point-like X-ray sources embedded in diffuse hot bubble X-ray emission. 
Since this shared trait required nuanced analysis, we consider these five nebulae here as a group. 
Based on the analysis in \S~\ref{embedded_section}, we conclude that NGC 2392 and NGC 6543 are significantly blended with hot bubble diffuse X-ray emission.
The original fits of the point-like component in these two nebulae (Table~\ref{specfitstable}) are apparently skewed by the hot bubble emission but, as described in \S~\ref{embedded_section}, we were able to spectroscopically isolate the point-like sources to better determine their X-ray properties.
The fixed-$N_H$ and free-$N_H$ spectral models for the isolated point-like components of NGC 2392 and NGC 6543 summarized in Table~\ref{embeddedsources} show general agreement in the $T_X$ and model normalization parameters but not in $N_H$, which is unconstrained in the free-$N_H$ model. 
This is not surprising as the high $T_X$ determined for the point-like components is less sensitive to absorption. 
Hence the $N_H$ best-fit value, or lack thereof, derived in the free-$N_H$ model carries little weight.
We therefore adopt the fixed-$N_H$ models for NGC 2392 and NGC 6543, which use the value of $N_H$ determined from $c_{H\beta}$ for these two nebulae.
Similarly, for the point-like component of NGC 7009, the model parameters (Table~\ref{specfitstable}) agree within errors, but again $N_H$ is poorly constrained in the free-$N_H$ model, so we adopt the fixed-$N_H$ model. 
For NGC 6826 the fixed-$N_H$ and free-$N_H$ models are inconsistent across all fit parameters.  
This is likely due to contamination by the diffuse component in NGC 6826, but it is difficult to isolate the point-like spectrum due to the low number of counts. 
We adopt the value of $N_H$ determined from $c_{H\beta}$ for NGC 6826, but urge caution in interpreting the derived X-ray properties.
There is no compelling evidence for close binary companions to any of these CSPNe, but all display P Cygni line profiles. 
The high plasma temperature of the compact source in NGC 2392 is similar to that of LoTr 5, suggesting a potential binary origin. 
The plasma temperature of the point source in NGC 2392 is the hottest measured from a central star thus far and there is growing indirect and direct evidence for a low mass binary companion to its CSPN  \citep{2012IAUS..282..470D,2014MNRAS.440.2684P}.  
We conclude that, for this group of CSPNe, wind shocks and/or heretofore unknown binary companions likely are responsible for the point-like X-ray emission.

\subsection{The central star of the Helix nebula}

As the brightest and best-studied example of ``hard'' X-ray emission from a CSPN, the CSPN of NGC 7293 remains in its own class. 
Our two models are in good agreement, so we adopt the free-$N_H$ model for the constraint on $N_H$. 
The origin of the hard ($T_X \sim 14$~MK) component in the Helix CSPN X-ray spectrum remains unknown \citep[see discussion in ][]{2012AJ....144...58K}. 
Late-type binary companions as cool as late M stars \citep{1999AJ....118..488C} appear to be ruled out and no wind has been measured from the central star \citep{2013A&A...553A.126G}.  
\citet{2001AJ....122..308G} report variable H$\alpha$ that is consistent with a dMe companion, but potential variations from additional lines may point towards another, as yet, unknown origin.
Infrared observations further constrain the existence of a cool companion and identify a dusty debris disk \citep{2007ApJ...657L..41S}. 
Active accretion of material from this debris disk onto the CSPN is unlikely given the large gap between the star and the inner edge of the debris disk \citep[$\sim35$ AU;][]{2007ApJ...657L..41S}. 
On the other hand, the lack of a strong stellar wind from the CSPN relaxes the conditions of nebular back flow \citep{2001MNRAS.328.1081S}. 
Furthermore, ionization of any backflowing material could give rise to the variable emission lines reported by \citet{2001AJ....122..308G}. 

Our spectral fit for NGC 7293 is very similar to that of \citet{2001ApJ...553L..55G}, but we note that, in these best-fit models, there are significant discrepancies between the model and data near 1 keV. 
These discrepancies may suggest the presence of multi-temperature components and/or emission lines \citep{2002ApJ...570..245S}. 
High-resolution grating X-ray spectroscopy of the central star could be used to identify the nature of this hard X-ray emitting source by resolving these potential spectral lines, thereby facilitating the study of the plasma diagnostics and/or search for asymmetries in line profiles from mass loss as demonstrated in such studies of O-stars \citep[e.g.,][]{2014MNRAS.439..908C}. 

\newpage
\subsection{Other CSPNe}

HBDS 1, NGC 1360, NGC 1514, NGC 4361, NGC 6445, and NGC 7008: 
For these CSPNe, the potential origin(s) of the X-ray emission remain essentially unconstrained. 
The spectral fit for NGC 7008 does not reproduce the soft X-ray spectral bin (see Figure~\ref{spectrafig}), suggesting the need for an alternative model and/or additional components. 
We adopt the fixed-$N_H$ model, but advise caution in interpreting the model-derived parameters for NGC 7008.
In HBDS 1, NGC 4361 and NGC 6445, the free-$N_H$ model leads to unconstrained $N_H$ and/or unphysical X-ray fluxes, so we adopt fixed-$N_H$.
In the case of NGC 1514 and NGC 1360, the free-$N_H$ models suggest additional column density provided by the metal-enriched nebulae themselves but the models are almost identical in other parameters. 
To avoid the uncertainty inherent in these free-$N_H$ models, we adopted the fixed-$N_H$ models for both of these PNe. 

Optical spectroscopy of the CSPN of NGC 1514 reveals an A-type companion spectrum, but the binary period remains uncertain \citep[see note in][]{2008PhDT.......109F}.
For the remaining sources, we only have speculative and/or insufficient evidence regarding the binarity. 
Morphologies for these PNe trend towards the more exotic, from the pair of rings in mid-IR imaging of NGC 1514 \citep{2010AJ....140.1882R} to spiral-like structures seen in NGC 4361 \citep{1999MNRAS.308..939V}. 
These morphologies are often used as indicators of binarity, but we reiterate that only for NGC 1514 does direct evidence exist for a binary central star. 
None of the sources listed above have measured wind properties, and UV spectral observations of NGC 1514, NGC 7008, and NGC 6445 were unable even to detect the stellar photospheres. 
One commonality among most of these PNe --- the exception being NGC 6445 --- is the lack of sharply-defined inner nebular rims. 
Such a characteristic makes these sources potential candidates for PN fallback.

\subsection{Variability in the X-ray observations}

X-ray variability is an important characteristic that can help distinguish the X-ray emission mechanism. 
For active companions we might expect flaring behavior in X-ray emission \citep{2005ApJS..160..511K}, while more rapidly varying emission is expected from accretion onto a compact CSPN.
Many of our observations are photon-starved, limiting our ability to detect rapid variability (i.e., variability on timescales $\stackrel{<}{\sim}1-10$ ks). 
We have, however, performed a search for flare-like variations in the 0.3-3.0 keV energy range of our point-like detections using the CIAO routine \verb+glvary+. 
This routine uses the Gregory-Loredo algorithm \citep{1992ApJ...398..146G} to search for non-constant flux throughout the observation\footnote{See \url{http://cxc.harvard.edu/csc/why/gregory\_loredo.html} for a description of the implementation of the Gregory-Loredo algorithm in CIAO and its use in the Chandra Source Catalog.}.
In most of our targets, the variability index of the Gregory-Loredo algorithm is 0 or 1 on a scale ranging from 0 for not variable with high probability to 10 for variable with a high probability. 
The highest variability index among our sample is reached in four of our targets (Lo 16, NGC 1360, NGC 4361, and NGC 6337), where the variability index is 2, which is also consistent with a low probability of variability. 
After examination of the light curves and high-energy background we conclude that, for these four sources, the variability index is higher because of the background behavior. 
We conclude that there is no evidence for X-ray variability on the part of any of the CSPN sources. 
More sensitive, longer-duration X-ray observations will be required to draw firmer conclusions in this regard. 

\begin{deluxetable*}{llcccccccccccc}
\tablewidth{0pt}
\tabletypesize{\scriptsize}
\tablecaption{Point-like Sources of X-ray Emission in Planetary Nebulae \label{cspnebasic}} 
\tablehead{
 \colhead{Object} & \colhead{$D$} & \colhead{$T_{\rm star}$}  & \colhead{log $L_{\rm bol}$} & \colhead{S.T.} & \colhead{Binary$^{a}$} & \colhead{Wind$^{b}$} & \colhead{Mo-} &  log $N_H$ & $T_{\rm X}$ & log $L_{\rm X}$ \\ 
 & (kpc) & (kK) &  ($L_{\odot}$) & & & & \colhead{del$^{c}$} & (cm$^{-2}$) & (MK) & (erg s$^{-1}$) } 
\startdata
DS 1                     & 0.73 & 90 & 3.48 & O(H) & c & -- & 1 & 20.88 & 14.72$_{-1.89}^{+12.20}$ & 29.83$_{-0.29}^{+0.49}$ \\
HBDS 1                   & 0.80 & 114 & 3.66 & O(H)    & (y)        & y & 1 &20.81 & 2.11$_{-0.15}^{+0.17}$ & 29.82$_{-0.27}^{+0.27}$ \\
HFG 1                    & 0.60 & 100 & 3.33 & O(H) & c & n & 1 &21.37 & 8.81$_{-1.61}^{+9.30}$ & 30.54$_{-0.27}^{+0.44}$ \\
LO 16                    & 0.84 & $\geq$82 & -- & O(H) & (c) & -- & 1 &21.47 & 15.87$_{-2.56}^{+14.24}$ & 29.13$_{-0.31}^{+0.47}$ \\
LOTR 5$^{\dagger}$                   & 0.50 & 100 & 2.01 & O(H) & (c) & y & 1 & 0.00 & 18.00$_{-4.00}^{+4.00}$ & 30.84$_{-0.05}^{+0.05}$ \\
NGC 246$^{\dagger\dagger}$                  & 0.50 & 140 & 3.63 & PG1159 & w        & y & 1 & 20.01 & 0.41$_{-0.00}^{+0.00}$: & 31.31$_{-0.27}^{+0.27}$: \\
NGC 1360                 & 0.38 & 110 & 3.30 & O(H) & (c)      & n & 1 & 19.53 & 1.39$_{-0.31}^{+0.31}$ & 29.41$_{-0.33}^{+0.72}$ \\
NGC 1514                 & 0.37 & 60 & 4$^{\dagger\dagger\dagger}$ & O(H) & (c)     & u & 1 & 21.44 & 1.79$_{-0.12}^{+0.20}$ & 29.64$_{-0.29}^{+0.29}$ \\
NGC 2371-72$^{\dagger\dagger}$              & 1.41 & 100 & 2.98 & [WO1] & --      & y & 1 & 20.38 & 0.32$_{-0.06}^{+0.00}$: & 31.76$_{-0.26}^{+3.11}$: \\
NGC 2392                 & 1.28 & 47 & 3.82 & Of(H) & w       & y & 1 & 20.90 & 36.28$_{-7.81}^{+19.64}$ & 30.58$_{-0.08}^{+0.08}$ \\
NGC 4361                 & 0.95 & 126 & 3.53 & O(H) & --    & n &  2 & $<$21.4 & 0.69$_{-0.04}^{+0.21}$ & 30.28$_{-0.33}^{+0.75}$ \\
NGC 6337                 & 0.86 & 105 & 2.6$^{\dagger\dagger\dagger}$ & --   & c & -- & 2 & 22.03$_{-0.10}^{+0.31}$ & 8.28$_{-4.78}^{+2.45}$ & 30.93$_{-0.33}^{+2.67}$ \\
NGC 6445                 & 1.39 & 170 & 2.97 & -- & --       & : & 1 &  21.64 & 10.71$_{-6.95}^{+3.39}$ & 30.48$_{-0.28}^{+0.64}$ \\
NGC 6543                 & 1.50 & 48 & 3.61 & Of-WR(H) & --      & y & 1 & 20.53 & 10.07$_{-1.77}^{+2.05}$ & 29.98$_{-0.11}^{+0.15}$ \\
NGC 6826                 & 1.30 & 50 & 3.81 & O3f(H) &--      & y & 1 & 20.71 & 2.17$_{-0.48}^{+0.39}$ & 30.32$_{-0.30}^{+0.38}$ \\
NGC 7008$^{\dagger\dagger}$                 & 0.70 & 97 & 3.12 & O(H) & w & n & 1 & 21.27 & 7.50$_{-4.90}^{+2.04}$: & 29.62$_{-0.30}^{+1.02}$: \\
NGC 7009                 & 1.45 & 87 & 3.67 & O(H) & --     & y & 1 &  20.61 & 2.85$_{-0.69}^{+2.54}$ & 30.41$_{-0.36}^{+0.37}$ \\
NGC 7094$^{\dagger\dagger}$                 & 1.39 & 110 & 3.61 & PG1159 & (y)   & y & 1 &  20.61 & 11.79$_{-2.54}^{+10.01}$: & 29.79$_{-0.32}^{+0.49}$: \\
NGC 7293                 & 0.22 & 110 & 1.95 & DAO & --      & n & 2 & 20.29$_{-0.32}^{+0.45}$ & 11.20$_{-0.30}^{+0.26}$ & 29.88$_{-0.26}^{+0.26}$ \\
SP 1                     & 1.13 & 72 & 3.44 & O(H)    & c & y & 1 & 21.42 & 3.17$_{-0.64}^{+3.18}$ & 29.90$_{-0.37}^{+0.36}$ \\
\enddata
\tablecomments{Source of supplementary CSPNe properties and adopted X-ray spectral model are discussed in \S\ref{resultssection}.}
\tablenotetext{a}{Binary notation scheme: `c' -- close binary, `w' -- wide binary, `y' -- binary with unknown period, `o' -- an optical pair; additionally, we indicate suspected binarity by placing the the binary indicator within parentheses.} 
\tablenotetext{b}{Stellar wind notation is from \citet{1985ApJ...291..237C}: `y' -- P Cygni profiles were detected, `n' -- no P Cygni profiles were detected and the stellar photosphere is detected, `:' -- the observation is not sensitive to the photosphere, `--' -- no UV satellite observations are present. }
\tablenotetext{c}{Adopted model: 1 -- $N_H$ determined from $c_{H\beta}$ (fixed-$N_H$ model), 2 -- $N_H$ determined from spectral fit (free-$N_H$ model).}
\tablenotetext{$\dagger$}{X-ray emission properties of LoTr 5 are adopted from \citet{2010ApJ...721.1820M}.}
\tablenotetext{$\dagger\dagger$}{Indicates spectral fits are unreliable for NGC 246, NGC 2371-72, NGC 7008, and NGC 7094, as discussed in \S\ref{resultssection}.} 
\tablenotetext{$\dagger\dagger\dagger$}{$L_{\rm bol}$ is estimated. For NGC 1514, $L_{\rm bol}$ is estimated from the spectral energy distribution presented by \citet{2010AJ....140.1882R} using the bolometric correction relationship from \citet{1984ApJ...279..304C}. For NGC 6337, $L_{\rm bol}$ is taken from \citet{1985Ap&SS.113...59A}, which is also the origin of the reported temperature.}
\end{deluxetable*}

\subsection{Non-Detections and Upper Limits\label{ulsection}}

For undetected sources, we estimate the $3\sigma$ detection threshold and $3\sigma$ background-subtracted count rates to estimate source detectability and X-ray upper limit fluxes and luminosities.
First, we determined the 0.3-3.0 keV background count rate in a source-free region near the undetected central star, with an extraction radius of 10$^{\prime\prime}$.  
We scaled these background counts to an extraction radius of 3$^{\prime\prime}$ and adopt this normalized count level as the total background counts, ${\rm CNT}_{\rm BG}$. 
The detection threshold, in counts, is the given by ${\rm CNT}_{\rm BG} + 3.0 \times \sqrt{{\rm CNT}_{\rm BG}}$. 
The background-subtracted count rate is given by ${\rm CR}_{\rm UL} = (3.0 \times \sqrt{{\rm CNT}_{\rm BG}})/T$, where $T$ is the total exposure time. 
We use this background-subtracted count rate to calculate the upper limit fluxes. 
The method described above can sometimes underestimate the limits for an undetected central star that is embedded in diffuse X-ray emission because the diffuse emission effectively raises the local background. 
For diffuse sources with no point-like components we used the following method to estimate the background count rate: we draw a series of events from a simulated PSF \citep[simulated by ChaRT, the {\it Chandra} Ray Tracer;][]{2003ASPC..295..477C}, add them to the diffuse events, and produce images with $0\farcs24$ pixels. 
Through visual inspection of these images we estimate the total number of PSF events required to produce an apparent point-like source against the diffuse emission. 
These estimates vary from source to source due to the brightness and compactness of the diffuse X-ray emission. 
The results of all of our upper limit estimates are included in Table~\ref{upperlimits}. 

To convert from count rates to observed (absorbed) and intrinsic (unabsorbed) X-ray fluxes, we input the background-subtracted upper limit count rates, ${\rm CR}_{\rm UL}$, into PIMMS simulations for a given observation cycle, absorption, and plasma temperature.
The $N_H$ values in our simulations are those derived from the $c_{H\beta}$ values for each PN. 
For consistency with our spectral fitting approach, we use an APEC thermal plasma with solar abundances and simulate two models, one with $T_X = 1$~MK and other other with $T_X = 8$~MK.
PIMMS then yields the observed and unabsorbed flux in the 0.3-3.0 keV range that corresponds to our input upper limits on count rate.
The results of these simulations are provided in Table~\ref{upperlimits}.
As expected, the upper limits for $T_X = 1$~MK are more sensitive to the assumed $N_H$, then upper limits for $T_X = 8$~MK. 
The X-ray flux upper limits are higher for $T_X = 1$~MK than for $T_X = 8$~MK.

\begin{deluxetable*}{lcccccccc}
\tablewidth{0pt}
\tabletypesize{\scriptsize}
\tablecaption{Upper Limits for Point-like Sources in  Planetary Nebulae \label{upperlimits}} 
\tablehead{
 \colhead{Object} & \colhead{${\rm CNT}_{\rm BG}$} & \colhead{$3\sigma {\rm CNT}_{\rm thresh}$} &  \colhead{$t_{\rm exp}$}  & \colhead{$CR_{\rm UL}$} & \colhead{log $N_H$}  & \colhead{log $F_{\rm X, UL_{\rm 1 MK}}$}  & \colhead{log $F_{\rm X, UL_{\rm 8 MK}}$} \\ 
 & (counts) & (counts) & (ks) &  (counts s$^{-1}$) & (cm$^{-2}$) & (erg cm$^{-2}$ s$^{-1}$) &   (erg cm$^{-2}$ s$^{-1}$)}
\startdata
Abell 33 & 0.93 & 3.82 & 29.7 & 9.75e-05 & 19.00 & -14.80 & -15.31 \\
Abell 65 & 1.02 & 4.06 & 29.7 & 1.02e-04 & 21.00 & -14.33 & -15.07 \\
BD+30 3639$^{\dagger}$ & -- & 300 & 19.0 & 1.58e-02 & 21.17 & -14.36 & -14.90 \\
HaWe 13 & 0.46 & 2.51 & 18.6 & 1.10e-04 & 21.39 & -13.73 & -14.76 \\
Hb 5 & 1.21 & 4.51 & 29.1 & 1.13e-04 & 21.78 & -12.60 & -14.21 \\
He 2-11 & 0.65 & 3.07 & 29.6 & 8.18e-05 & 21.84 & -12.50 & -14.24 \\
IC 418$^{\dagger}$  & -- & 15 & 49.4 & 3.04e-04 & 21.00 & -14.55 & -15.19 \\
IC 1295 & 1.02 & 4.06 & 29.7 & 1.02e-04 & 21.26 & -14.01 & -14.91 \\
IC 2149 & 0.56 & 2.80 & 28.7 & 7.81e-05 & 21.09 & -14.34 & -15.14 \\
IC 4637 & 1.02 & 4.06 & 29.6 & 1.03e-04 & 21.48 & -13.58 & -14.70 \\
IC 5148/50 & 0.74 & 3.33 & 19.7 & 1.31e-04 & 19.53 & -14.64 & -15.17 \\
M 1-26 & 1.21 & 4.51 & 26.7 & 1.23e-04 & 21.74 & -12.71 & -14.25 \\
M 1-41 & 0.74 & 3.33 & 29.6 & 8.74e-05 & 21.88 & -12.33 & -14.13 \\
M 27 & 1.12 & 4.28 & 19.8 & 1.60e-04 & 20.38 & -14.48 & -15.04 \\
M 57 & 0.65 & 3.07 & 19.8 & 1.22e-04 & 20.84 & -14.40 & -15.06 \\
M 76 & 1.12 & 4.28 & 29.9 & 1.06e-04 & 20.71 & -14.53 & -15.16 \\
M 97 & 0.74 & 3.33 & 19.2 & 1.34e-04 & 19.83 & -14.63 & -15.16 \\
NGC 40 & 0.93 & 3.82 & 19.9 & 1.45e-04 & 21.33 & -14.11 & -14.88 \\
NGC 1501 & 0.46 & 2.51 & 19.7 & 1.04e-04 & 21.53 & -13.45 & -14.64 \\
NGC 2346 & 0.74 & 3.33 & 29.9 & 8.66e-05 & 21.41 & -13.80 & -14.85 \\
NGC 2438 & 1.30 & 4.72 & 29.7 & 1.15e-04 & 20.97 & -14.32 & -15.04 \\
NGC 2899 & 1.12 & 4.28 & 29.6 & 1.07e-04 & 21.33 & -13.85 & -14.83 \\
NGC 3132 & 0.65 & 3.07 & 24.0 & 1.01e-04 & 20.71 & -14.98 & -15.40 \\
NGC 3242$^{\dagger}$  & -- & 20 & 29.3 & 6.83e-04 & 20.31 & -14.63 & -15.19 \\
NGC 3918$^{\dagger}$  & 0.65 & 3.07 & 30.0 & 8.05e-05 & 21.12 & -14.30 & -15.11 \\
NGC 6026 & 1.21 & 4.51 & 29.6 & 1.11e-04 & 21.42 & -13.67 & -14.73 \\
NGC 6072 & 0.74 & 3.33 & 29.6 & 8.75e-05 & 21.48 & -13.63 & -14.77 \\
NGC 6153 & 0.65 & 3.07 & 28.6 & 8.46e-05 & 21.67 & -13.12 & -14.53 \\
NGC 6302 & 0.56 & 2.80 & 22.5 & 9.94e-05 & 21.66 & -13.09 & -14.47 \\
NGC 6369$^{\dagger}$  & 1.21 & 4.51 & 29.6 & 1.11e-04 & 21.86 & -12.31 & -14.07 \\
NGC 6772 & 0.84 & 3.58 & 29.4 & 9.34e-05 & 21.51 & -13.54 & -14.71 \\
NGC 6781 & 0.74 & 3.33 & 28.3 & 9.16e-05 & 21.43 & -13.74 & -14.81 \\
NGC 6804 & 1.39 & 4.94 & 29.6 & 1.20e-04 & 21.45 & -13.59 & -14.67 \\
NGC 6894 & 0.65 & 3.07 & 29.2 & 8.27e-05 & 21.50 & -13.61 & -14.77 \\
NGC 7027$^{\dagger}$  & -- & 40 & 18.2 & 2.20e-03 & 21.69 & -13.23 & -14.43 \\
NGC 7076 & 1.02 & 4.06 & 29.6 & 1.03e-04 & 21.62 & -13.18 & -14.51 \\
NGC 7354 & 0.56 & 2.80 & 29.6 & 7.57e-05 & 21.81 & -12.67 & -14.33 \\
NGC 7662$^{\dagger}$  & -- & 20 & 27.6 & 7.25e-04 & 20.74 & -14.55 & -15.18 \\
Sh 2-71 & 1.12 & 4.28 & 29.6 & 1.07e-04 & 21.72 & -12.85 & -14.34 \\
\enddata
\vspace{-0.4cm}
\tablecomments{
Upper limit calculations are described in \S\ref{ulsection} and briefly described here. 
${\rm CNT}_{\rm BG}$ gives the background counts scaled to the extraction region of $3^{\prime\prime}$, $3\sigma {\rm CNT}_{\rm thresh}$ gives the count rate required from a signal that is $3\sigma$ above the background, and $CR_{\rm UL}$ gives the background-subtracted $3\sigma$ count rate that is used to estimate the upper limit X-ray fluxes.
We used the $N_H$ determined from $c_{H\beta}$ measurements and two plasma temperatures of 1 MK and 8 MK to calculate intrinsic source fluxes, $F_{\rm X, UL_{\rm 1 MK}}$ and $F_{\rm X, UL_{\rm 8 MK}}$, respectively.
}
\tablenotetext{$\dagger$}{Indicates that the presence of diffuse emission led to an overestimate and we adopted  $3\sigma {\rm CNT}_{\rm thresh}$  using ChaRT/MARX simulations described in the text. }
\end{deluxetable*}

\begin{deluxetable*}{llcccccccccccc}
\tablewidth{0pt}
\tabletypesize{\scriptsize}
\tablecaption{Non-Detections of Point-like Emission in Planetary Nebulae \label{ulbasic}} 
\tablehead{
 \colhead{Object} & \colhead{$D$} & \colhead{$T_{\rm star}$}  & \colhead{log $L_{\rm bol}$} & \colhead{S.T.} & \colhead{Binary} & \colhead{Wind} & log $N_H$ & log $L_{\rm X,UL_{\rm 1 MK}}$ & log $L_{\rm X, UL_{\rm 8 MK}}$  \\ 
  & (kpc) & (kK) &  ($L_{\odot}$) & & & & (cm$^{-2}$) &  (erg s$^{-1}$) & (erg s$^{-1}$) } 
\startdata
Abell 33 & 1.16 & 100 & 2.27 & O(H) & w & n & 19.00 & 29.41 & 28.89 \\
Abell 65 & 1.17 & 83 & -- & CB & c & u & 21.00 & 29.89 & 29.14 \\
BD+30 3639 & 1.30 & 32 & 3.63 & [WC9] & --        & y & 21.17 & 31.01 & 30.92 \\
HaWe 13 & 1.01 & 68 & 1.95 & hgO(H) & (y)      & -- & 21.39 & 30.36 & 29.33 \\
Hb 5 & 1.70 & 172 & 3.64 & -- & --      & -- & 21.78 & 31.94 & 30.32 \\
He 2-11 & 1.14 & 90 & -- & -- & c & -- & 21.84 & 31.69 & 29.95 \\
IC 418 & 1.20 & 38 & 3.72 & Of(H) & --    & y & 21.00 & 29.22 & 29.14 \\
IC 1295 & 1.23 & 98 & 2.40 & hgO(H) & -- & -- & 21.26 & 30.25 & 29.34 \\
IC 2149 & 1.52 & 42 & 3.66 & Of(H) & --    & y & 21.09 & 30.10 & 29.30 \\
IC 4637 & 1.30 & 50 & 3.60 & O(H) & (o) & -- & 21.48 & 30.73 & 29.60 \\
IC 5148/50 & 0.85 & 110 & 2.07 & hgO(H) & -- & n & 19.53 & 29.29 & 28.77 \\
M 1-26 & 1.20 & 33 & 3.65 & Of(H) & --     & -- & 21.74 & 31.52 & 29.99 \\
M 1-41 & 1.47 & 187 & -- & -- & --  & -- & 21.88 & 32.09 & 30.28 \\
M 27 & 0.38 & 135 & 2.49 & DAO & (y) & : & 20.38 & 28.76 & 28.19 \\
M 57 & 0.70 & 148 & 2.58 & hgO(H) & -- & n & 20.84 & 29.37 & 28.70 \\
M 76 & 1.20 & 140 & 2.23 & PG1159 & o & -- & 20.71 & 29.70 & 29.07 \\
M 97 & 0.76 & 105 & 1.95 & hgO(H) & --  & n & 19.83 & 29.20 & 28.68 \\
NGC 40 & 1.02 & 45 & 3.60 & [WC8] & --   & y & 21.33 & 29.98 & 29.21 \\
NGC 1501 & 0.72 & 135 & 3.66 & [WC4] & (y)  & -- & 21.53 & 30.34 & 29.15 \\
NGC 2346 & 0.90 & 80 & -- & O(H)? & c & n & 21.41 & 30.18 & 29.13 \\
NGC 2438 & 1.42 & 124 & 2.31 & hgO(H) & --  & -- & 20.97 & 30.06 & 29.34 \\
NGC 2899 & 1.37 & 270 & -- & -- & y  & -- & 21.33 & 30.50 & 29.52 \\
NGC 3132 & 0.81 & 100 & 2.19 & -- & w & : & 20.71 & 28.91 & 28.50 \\
NGC 3242 & 1.00 & 89 & 3.54 & O(H) & --  & y & 20.31 & 29.42 & 29.33 \\
NGC 3918 & 1.84 & 150 & 3.70 & O(H) & --  & : & 21.12 & 30.30 & 29.50 \\
NGC 6026 & 1.31 & 35 & -- & -- & c & n & 21.42 & 30.64 & 29.58 \\
NGC 6072 & 1.39 & 140 & 2.59 & -- & --  & -- & 21.48 & 30.73 & 29.60 \\
NGC 6153 & 1.10 & 109 & 3.63 & wels & -- & -- & 21.67 & 31.04 & 29.63 \\
NGC 6302 & 1.17 & 220 & -- & -- & -- & : & 21.66 & 31.13 & 29.74 \\
NGC 6369 & 1.55 & 66 & 4.07 & [WO3] & (y)   & -- & 21.86 & 32.14 & 30.38 \\
NGC 6772 & 1.20 & 135 & 2.41 & -- & --  & -- & 21.51 & 30.69 & 29.53 \\
NGC 6781 & 0.95 & 112 & 2.55 & DAO & -- & -- & 21.43 & 30.29 & 29.22 \\
NGC 6804 & 1.47 & 85 & 3.71 & O(H) & (y) & -- & 21.45 & 30.82 & 29.74 \\
NGC 6894 & 1.31 & 100 & 2.23 & -- & --  & -- & 21.50 & 30.70 & 29.55 \\
NGC 7027 & 0.89 & 175 & 3.87 & -- & --  & -- & 21.69 & 29.82 & 29.74 \\
NGC 7076 & 1.47 & 80 & 3.28 & -- & -- & -- & 21.62 & 31.23 & 29.90 \\
NGC 7354 & 1.60 & 96 & 3.95 & -- & --  & -- & 21.81 & 31.82 & 30.15 \\
NGC 7662 & 1.26 & 111 & 3.42 & O(H) & --  & y & 20.74 & 29.64 & 29.56 \\
Sh 2-71 & 1.14 & 157 & 2.85 & -- & -- & -- & 21.72 & 31.34 & 29.85 \\
\enddata
\tablecomments{
Columns and description as in Table~\ref{cspnebasic}, but with X-ray upper limit $L_X$ calculated from Table~\ref{upperlimits}.
}
\end{deluxetable*}

\section{Discussion}\label{discussion}

We now discuss the spectral energy distributions of point-like sources of X-ray emission from the CSPNe observed thus far as part of the ChanPlaNS project. 
We limit discussion to the X-ray properties derived assuming an absorbed optically-thin thermal plasma model in the 0.3-3.0 keV energy range, as presented in \S\ref{resultssection}.
The detected point-like sources of X-ray emission (20 objects described in Table~\ref{cspnebasic}) and the undetected sources (39 objects described in Table~\ref{ulbasic}) comprise the sample to date. 
In the following, we highlight the insight provided by this sample of CSPNe and point towards future directions and studies made possible by  these observations.

\subsection{Evidence for metal-enhanced absorption}

There is some evidence in Figure~\ref{nhfits} for enhanced absorption to the X-ray emitting sources as compared to the absorption determined from the optical extinction to the nebula ($c_{H\beta}$). 
The enhancement is likely due to the metal-enriched nebular material absorbing some of the X-ray photons as they pass through. 
However, our limited sensitivity to low-energy photons, and the faintness of the sources, restrict our ability to rigorously constrain the absorption. 
Future X-ray observations with high sensitivity and high spectral resolution in the 0.3-1.0 keV range may potentially resolve the absorption edges of the nebular metals in the spectra of X-ray bright CSPNe. 

\subsection{Tracing the origin of X-ray emitting CSPNe}
 
We compared the physical properties of the sample of CSPNe to their X-ray emission properties to search for the origin of the X-ray emission. 
In Figure~\ref{hrdiagram}, we present the approximate locations of the CSPNe on the H-R diagram. 
From Figure~\ref{hrdiagram}, we infer that a larger fraction ($50\%$) of high-luminosity CSPNe, here defined as having log~$L_{\rm bol}/L_{\odot}\stackrel{>}{\sim} 3$ including the borderline cases NGC 2371-72 and NGC 6445 (symbols 9 and 13, respectively, in Figure~\ref{hrdiagram}), are detected as X-ray emitting sources, compared to only 12\% of the lower-luminosity sources.
This suggests that the presence of X-ray emission might be correlated with the central star luminosity. 
Indeed, Figures~\ref{lxlbol}~\&~\ref{ratios} suggest that most of the X-ray emitting CSPNe with $L_{\rm bol}$ measurements or estimates have $L_{\rm X}/L_{\rm bol}$ ratios close to $10^{-7}$.
Figure~\ref{lxlbol} also reveals the sobering reality of the inability of our existing observations to rule out emission at this level from a majority of our undetected sources (we discuss this further below).

\begin{figure} 
\includegraphics[scale=0.55]{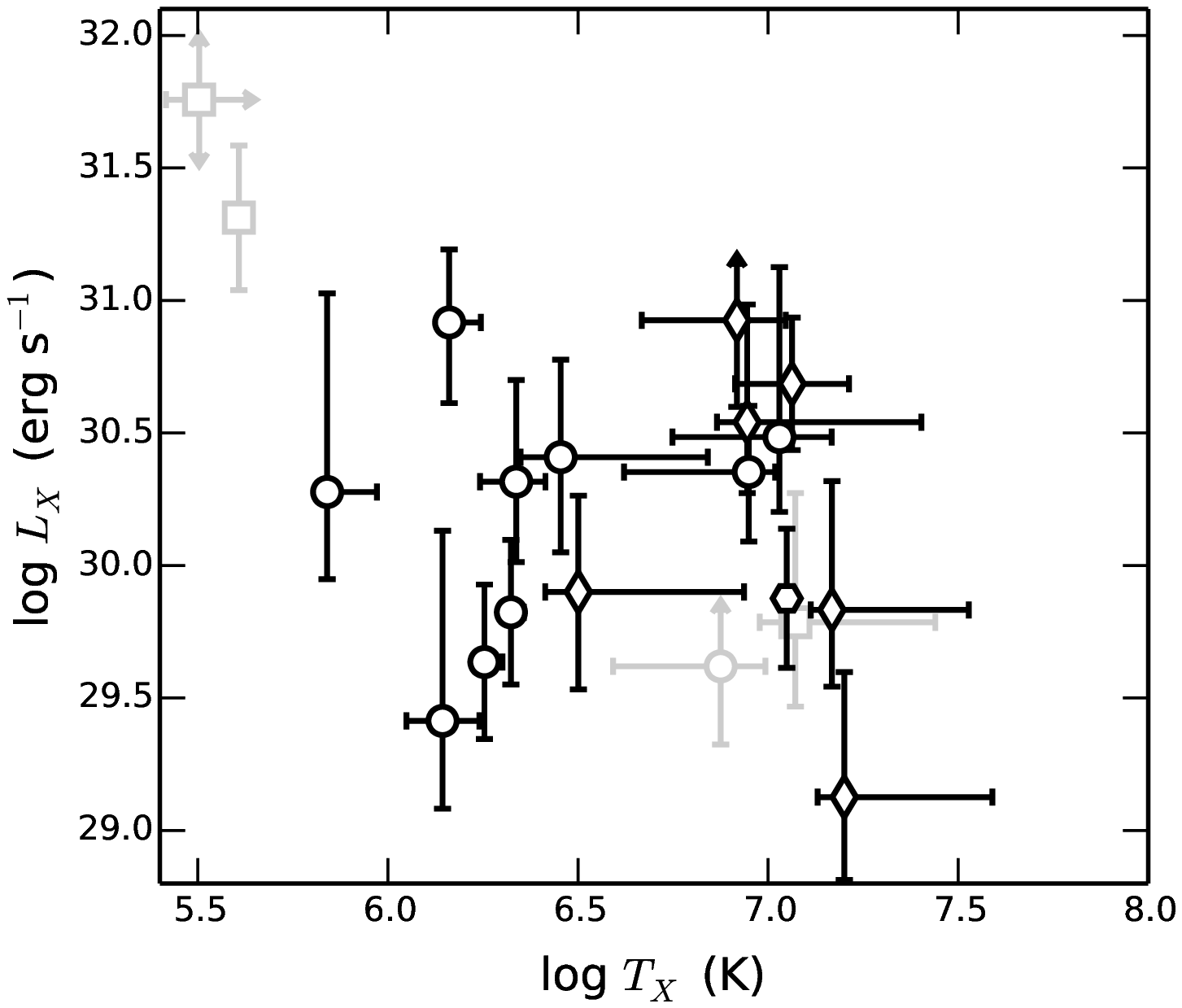}
\includegraphics[scale=0.55]{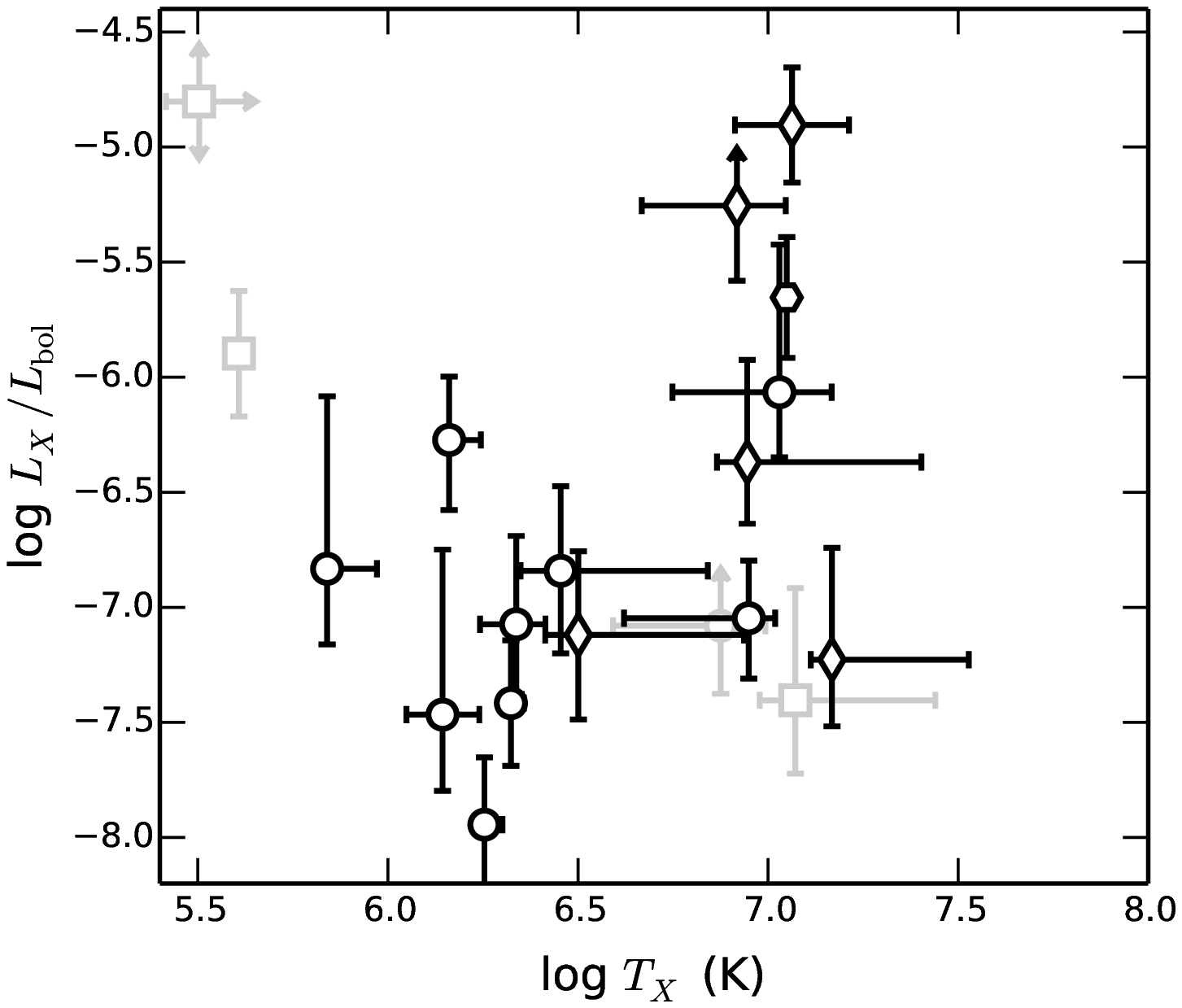}
\caption{X-ray characteristics of CSPNe obtained from the adopted models described in the text (see \S3).
Known binary CSPNe, hot CSPNe with WR or PG1159 spectral type, and the Helix CSPN have been identified as in Figure~\ref{medenhfig}. 
CSPNe with unreliable fits (NGC 246, NGC 2371-72, NGC 7008, \& NGC 7094) are shown as lighter colored symbols. 
\label{adoptedfits} }
\end{figure}

Figure~\ref{ratios} indicates that $\sim60\%$ of the X-ray emitting point-like sources have log~$L_{\rm X}/L_{\rm bol}\sim-7$. 
This fractional X-ray luminosity is too large to originate from the CSPNe photosphere (see Appendix), and is instead similar to that seen in X-ray emission from self-shocking stellar winds of O stars \citep[e.g.,][]{2011ApJS..194....7N}. 
This comparison suggests that a similar mechanism, i.e., self-shocking winds, might be producing the point-like X-ray emission from many, perhaps most, CSPNe.
This is further supported by the fact that during the high-luminosity phase of post-AGB evolution, where a majority of our detections are made, stellar winds are strongest and subsequently fade after the temperature of the central star has peaked \citep{1985ApJ...291..237C}.
Further evidence for self-shocking winds comes from UV spectroscopy where CSPNe with variable P Cygni profiles are observed with high ionization potential ions (e.g., \ion{O}{6}) despite low effective temperatures, which is interpreted as Auger ionization from X-ray emission due to shocks in their stellar winds \citep{2013A&A...553A.126G}.
For X-ray generation via self-shocking winds, the variation in the wind speed is more important than the absolute value. 
Variations of only a few hundred km~s$^{-1}$, or $\sim10\%$ of the typical wind speed measured from CSPN, are capable of producing X-ray emitting shocks.

It is tempting to conclude that all sources with log~$L_{\rm X}/L_{\rm bol}\sim-7$ are caused by self-shocking stellar winds. 
However, the CSPN of NGC 1360 lacks a strong stellar wind --- as determined from ultraviolet satellite spectroscopy --- and the CSPN of DS 1, which harbors a close binary system and whose companion  is likely responsible for the X-ray emission \citep{2010ApJ...721.1820M}, may only fall into this range coincidentally. 
The extant UV observations of NGC 1360 and other CSPNe without detected UV winds may not have been sensitive to weak P Cygni line profiles, but this should also make detection of the X-ray emission from wind shocks more difficult.
There is also the remote possibility that the lower ionization stages that would appear in UV emission are unpopulated because the shock-heated circumstellar atoms are at higher ionization stages, producing emission primarily in the X-ray regime. 
Though beyond the scope of this paper, a more rigorous comparison between the UV and X-ray emission of such metal-rich self-shocking stellar winds is clearly necessary. 

\begin{figure}
\includegraphics[scale=0.55]{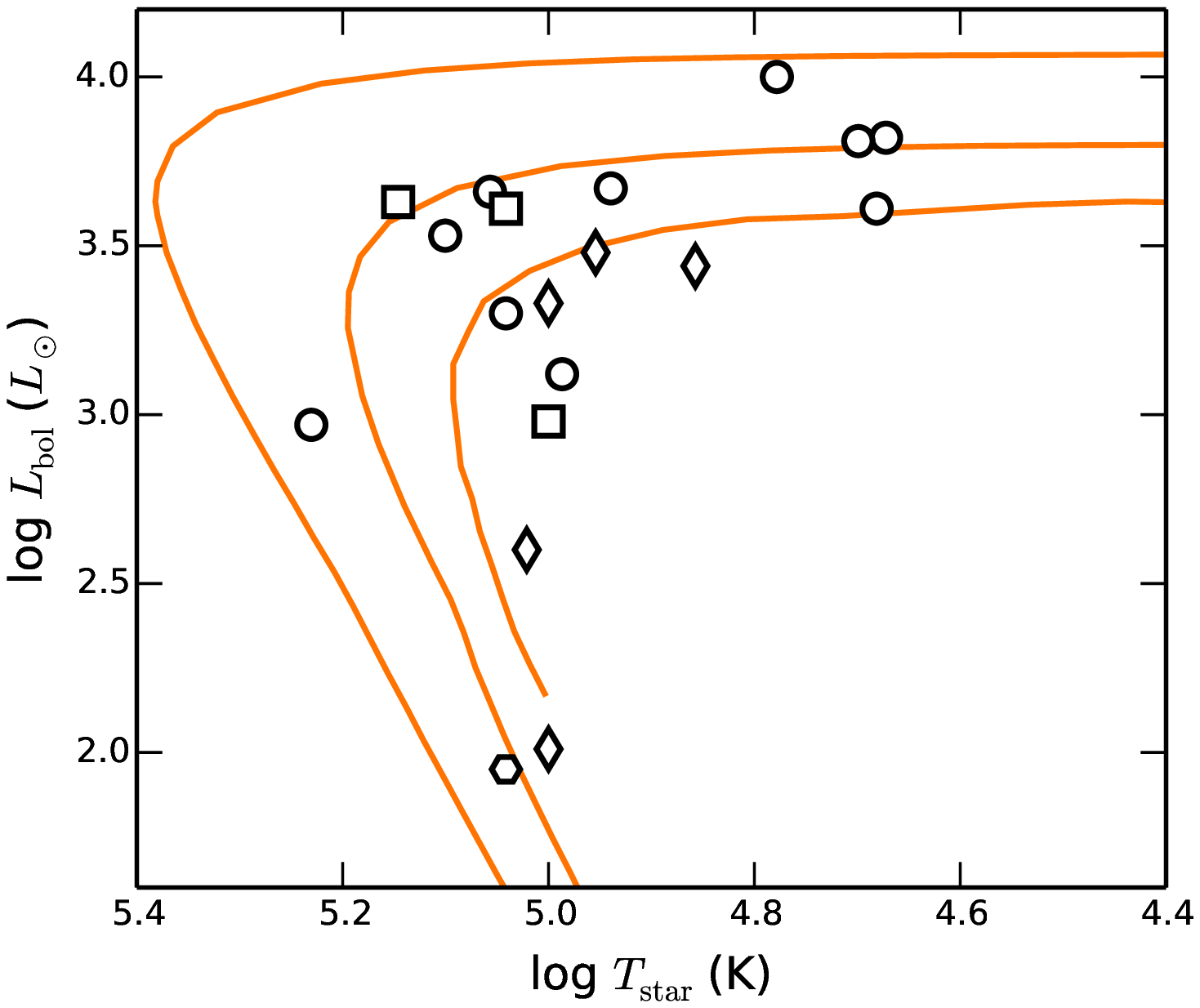}
\includegraphics[scale=0.55]{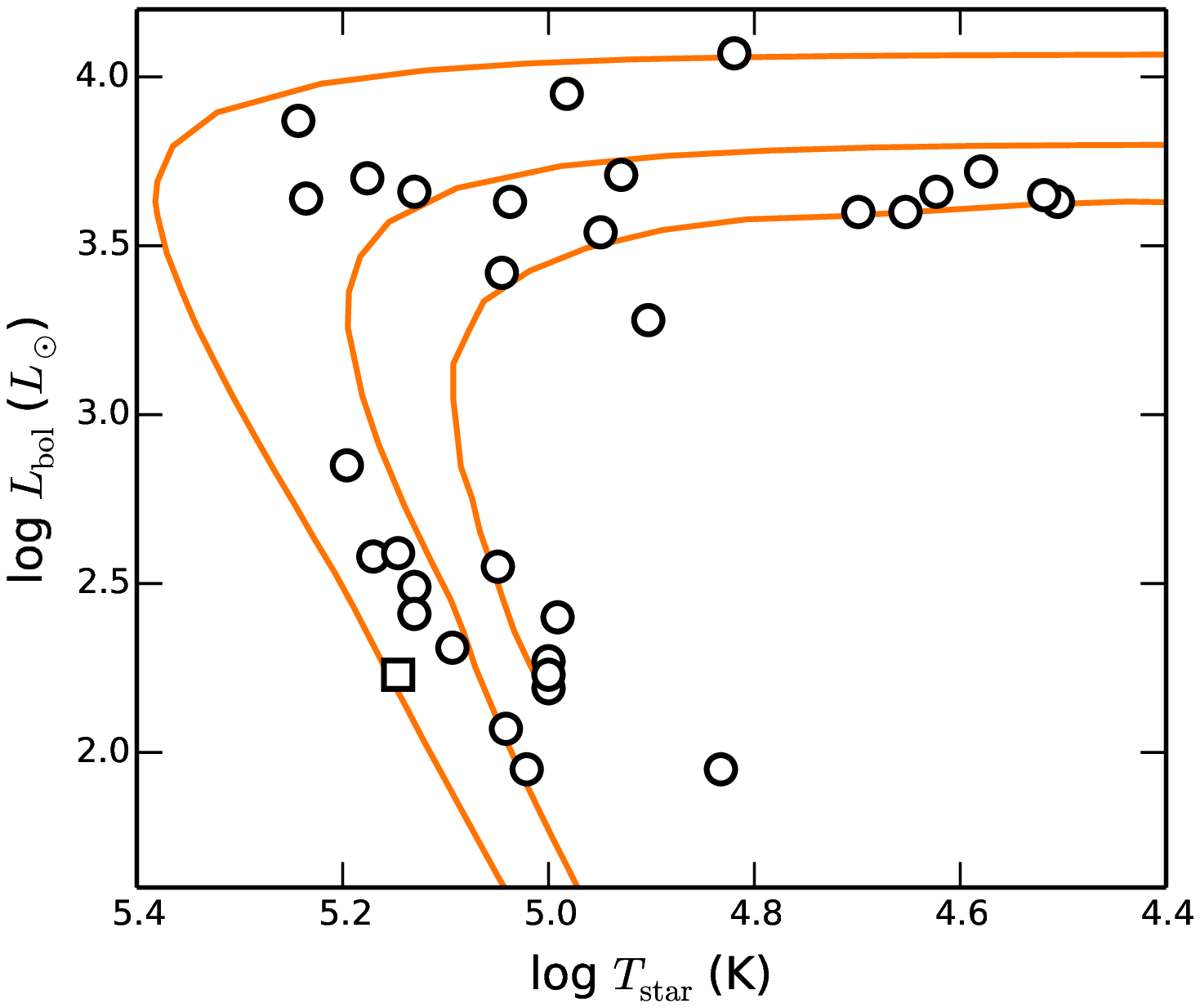}
\caption{
Approximate positions of our CSPNe on the H-R diagram, with post-AGB evolutionary tracks from \citet{1983ApJ...272..708S} and \citet{1995A&A...299..755B} overlaid. 
Detected sources are indicated on the left while non-detections are indicated on the right. 
Known binary CSPNe, hot CSPNe with WR or PG1159 spectral type, and the Helix CSPN have been identified as in Figure~\ref{medenhfig}. 
CSPNe without estimates of $L_{\rm bol}$ are not included in this plot. 
\label{hrdiagram} }
\end{figure}

\begin{figure}
\includegraphics[scale=0.55]{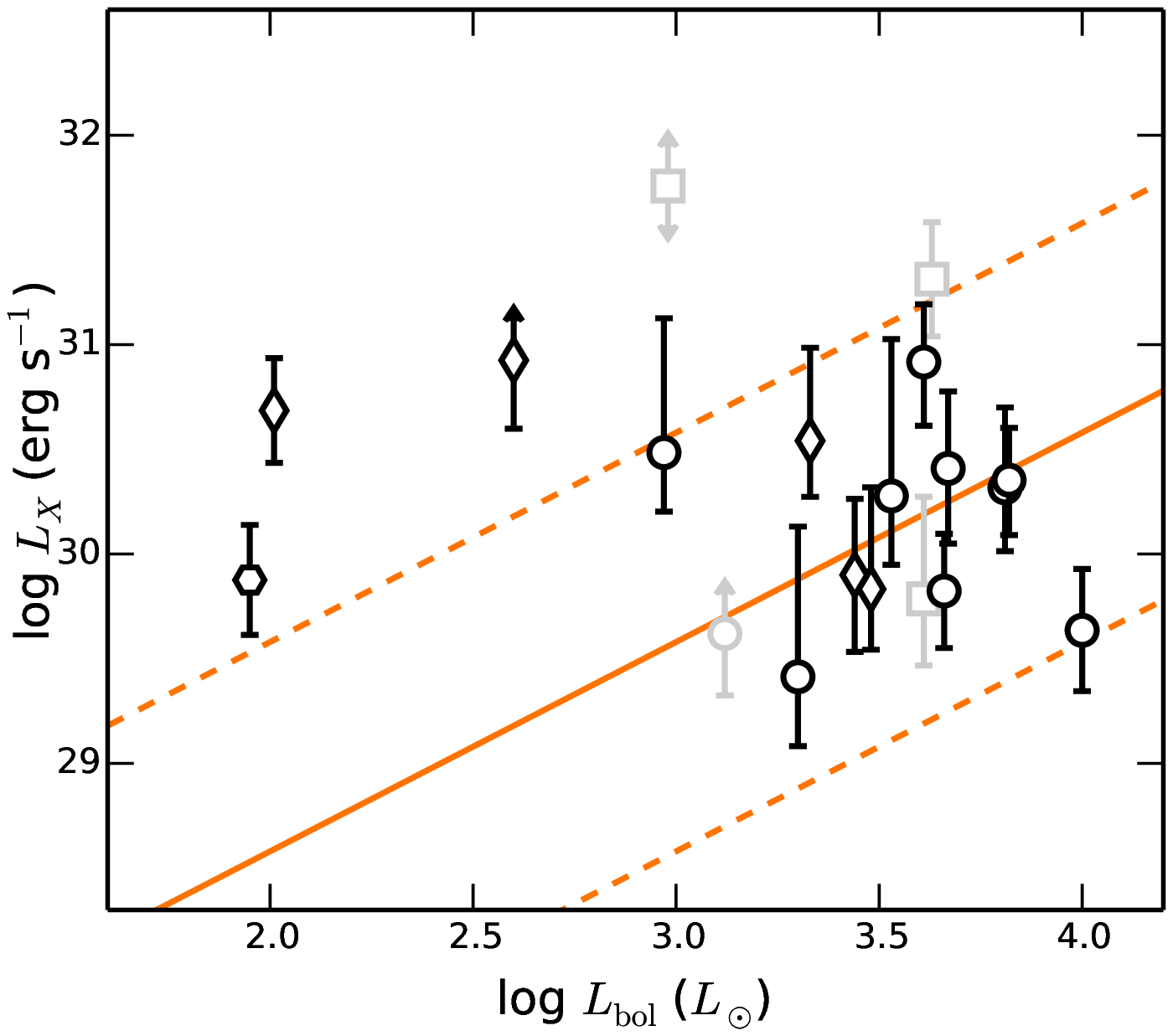}
\includegraphics[scale=0.55]{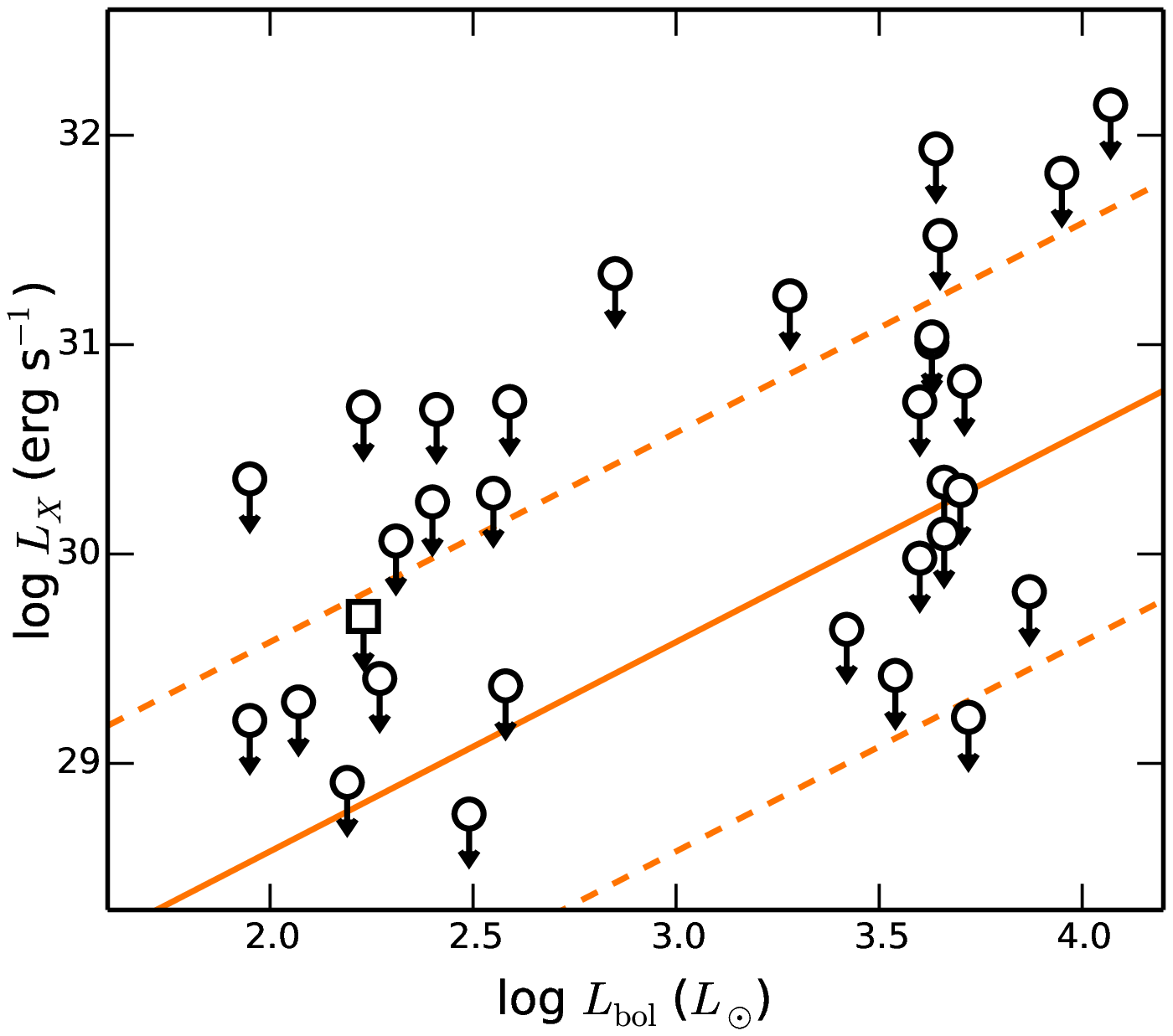}
\caption{
X-ray luminosity as determined from the adopted model best-fit parameters (left; Table~\ref{cspnebasic}) or upper limit estimates for a 1 MK thermal plasma (right; Table~\ref{ulbasic}) and appropriate distances versus the CSPN bolometric luminosity. 
Known binary CSPNe, hot CSPNe with WR or PG1159 spectral type, and the Helix CSPN have been identified as in Figure~\ref{medenhfig}. 
CSPNe with unreliable fits are shown as lighter colored symbols. 
In both panels, the overlaid lines illustrate log~$L_{\rm X}/L_{\rm bol}$ ratios of -8 (lowest dashed line), -7 (solid line), and -6 (highest dashed line).
\label{lxlbol} }
\end{figure}

\begin{figure}
\center 
\includegraphics[scale=0.55]{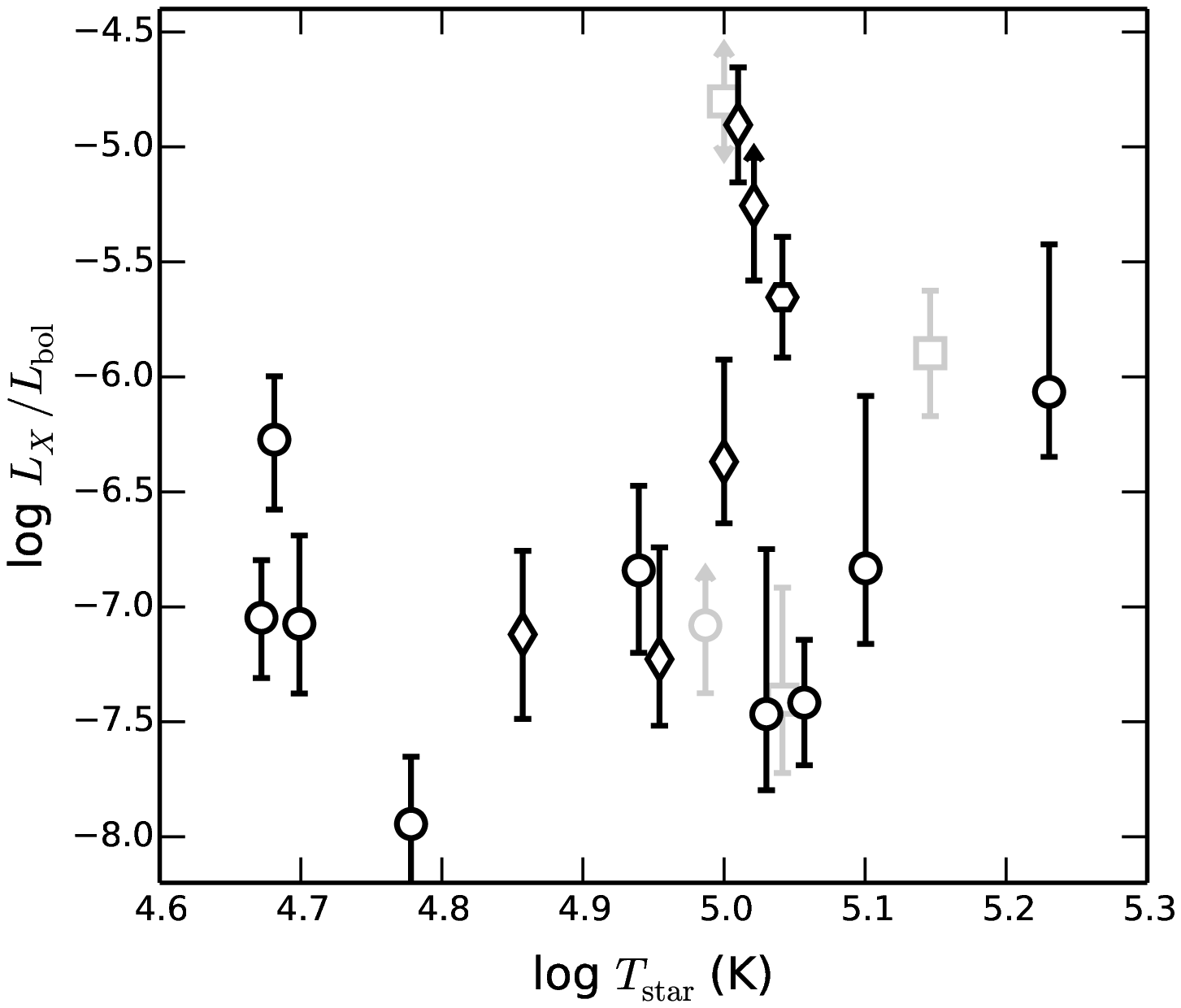}
\caption{\label{ratios} 
X-ray luminosity as determined from the best-fit adopted model scaled by bolometric luminosity of the CSPN versus CSPN temperature. 
Known binary CSPNe, hot CSPNe with WR or PG1159 spectral type, and the Helix CSPN have been identified as in Figure~\ref{medenhfig}. 
CSPNe with unreliable fits are shown as lighter colored symbols. 
Lo 16 does not appear on this plot as its $L_{\rm bol}$ is unknown.
} 
\end{figure}

The $L_{\rm X}/L_{\rm bol}$ ratios that deviate from log~$L_{\rm X}/L_{\rm bol}\sim-7$ are likely comprised of two different populations: CSPNe with binary companions, and CSPNe of the hot PG1159 spectral type. 
The CSPNe that are known or suspected binaries are spread throughout the $L_{\rm X}/L_{\rm bol}$ range,  consistent with their X-ray properties being uncorrelated with the CSPN properties. 
Indeed, several of these sources have been previously analyzed \citep{2010ApJ...721.1820M}, yielding evidence that a saturated (log~$L_{\rm X}/L_{\rm bol}\sim-3$) late-type binary companion is the source of the X-ray emission. 
The present study supports these previous results, in that the known binary CSPNe are among the hottest X-ray sources, with plasma temperatures near or above $\sim10^{7}$ K --- consistent with the X-rays arising in coronae associated with late-type binary companions. 
Nearly 70\% of such high-$T_X$ sources --- which represent the leftmost sources in Figure~\ref{medenhfig} --- have close binary systems or active companions. 
This trend towards hotter plasma temperatures amongst the close binary systems naturally suggests that the CSPNe of NGC 2392, NGC 6445, and NGC 7008 are ideal candidates for follow-up searches for heretofore undetected binary companions. 
Indeed, \citet{2014MNRAS.440.2684P} report a tentative close binary signature in the radial velocities measured for the CSPN of NGC 2392. 

Meanwhile, although the spectral fits are uncertain, the hot, hydrogen-deficient PG1159 stars trend towards higher $L_{\rm X}/L_{\rm bol}$ ratios, suggesting a contribution from photospheric emission, perhaps including NLTE effects, in addition to emission from self-shocking winds. 
\citet{2007ApJ...670..442H} found that the X-ray spectrum NGC 246 includes emission from the photospheric continuum but is dominated by a complex of \ion{C}{5} and \ion{C}{6} emission lines from 0.3-0.4 keV; such behavior is not well-modeled by our choice of plasma model, leading to a poor spectral fit and, hence, large uncertainty in the X-ray luminosity. 
Similar spectral characteristics are seen in the PG1159 CSPN of K 1-16, for which \citet{2013ApJ...766...26M} used a composite photospheric and carbon-enriched plasma model for the X-ray spectrum. 

\subsection{Alternative mechanisms}

The presence of stellar winds and binary companions amongst the X-ray emitting CSPNe naturally guides our attempt to find a causal relationship to the X-ray emission. 
But there are additional potential origins for CSPN X-ray emission. 
The magnetic wind confinement model (MWCM) has been offered as a potential explanation for the anomalously hard X-ray emission from certain young O stars; in the case of the O-star $\theta^{1}~{\rm Orionis~C}$, a fossil kG magnetic field is invoked to redirect the stellar wind towards the equator, where the northern and southern components collide \citep{2005ApJ...628..986G}. 
Although the X-ray emission from such a MWCM source would share characteristics with that of X-ray emission from CSPNe, we point out that kG large-scale magnetic fields have been ruled out from all CSPNe measured in a recent study by \citet{2014A&A...563A..43L}. 
Nine objects in the \citet{2014A&A...563A..43L} study are X-ray emitting CSPNe. 
A scaled down version of MWCM, with magnetic fields of a few hundred G, remains possible.
The value $L_{\rm X}/L_{\rm bol}\sim 10^{-7}$ that is characteristic of O-stars is not uniquely associated to self-shocking winds; in fact, colliding wind binaries (CWB) also display a similar ratio \citep{2011ApJS..194....7N}. 
In this case, stellar winds from two massive stars collide to form X-ray emitting plasma between the two stars. 
Such a scenario in CSPNe would require two stars driving stellar winds of similar strength yet no such candidates exist among the X-ray emitting CSPNe. 
Nevertheless, a search for synchrotron radiation from these X-ray emitting CSPNe may help determine if a CWB-mechanism is responsible for the X-ray emission. 

Accretion onto the CSPN is another potential origin for the observed X-ray emission; indeed, accretion disk winds have been invoked as a PN shaping mechanism \citep{2001ApJ...546..288B}.
In \citet{2010ApJ...721.1820M} we considered accretion in detail for the PCEB CSPNe DS 1, HFG 1, and LoTr 5 and found this scenario difficult to reconcile with observations. 
Many of the arguments made in \citet{2010ApJ...721.1820M} apply to the larger ChanPlaNS sample of X-ray emitting CSPNe. 
First, the maximum temperature for an accretion disk around a typical CSPN barely reaches $10^{6}$~K in the most compact CSPN, but these are also the most evolved CSPNe and are seldom detected in X-rays. 
In those CSPNe that are detected, the X-ray plasma temperature is an order of magnitude hotter than expected from an accretion disk.
Second, accretion directly onto the surface of a CSPN is possible but requires a reservoir for the accreting gas and dust, such as fallback of nebular material \citep{2001MNRAS.328.1081S,2009NewA...14..654F}.
Any spherical accretion onto the surface of the CSPN will have to contend with the outflowing stellar wind; thus spherical accretion is unlikely in CSPNe with strong winds, hence accretion, if it does occur, is likely to involve an accretion disk. 
To first approximation, the temperature of material free-falling onto a CSPN could reach $10^{8}$~K \citep{2002apa..book.....F}, but perhaps an isobaric cooling flow, such as is thought to arise from the boundary layer in WDs in cataclysmic variable (CV) systems, is more likely \citep{2005ApJ...626..396P}. 
Indeed, the X-ray characteristics of CSPN are similar to those observed from cooling flows in CVs, but with even cooler temperatures that could perhaps be attributed to the larger radii of the contracting CSPNe.
On the subject of accretion disks, pulsed jets launched by the young stellar objects may produce X-ray emission that remains stationary, at the base of the jets, and display X-ray properties similar to these observed from CSPNe \citep{2011ApJ...737...54B}.
Such a pulsed jet could be the shaping agent of PN. 
\citet{2014MNRAS.440.2684P} have found evidence for polar-oriented, high-velocity outflow in UV spectra of NGC 2392. 
If such polar outflows are ubiquitous features of CSPNe, shocks at the bases of these outflows should be considered among the potential origins of CSPN X-ray emission. 

\subsection{Detectability and the Undetected CSPNe} 

The nondetections and their estimated upper limits reveal the limitations of our observations in detecting the different classes of CSPN X-ray emission found thus far. 
For the most conservative upper limit estimates --- which are based on a plasma temperature of 1 MK, i.e., what we expect for wind shocks, our upper limits on $L_{\rm X}/L_{\rm bol}$ ratios suggest that the observations cannot rule out X-ray emission at a level with log~$L_{\rm X}/L_{\rm bol}\sim-7$ for a majority of our CSPNe (see Figure~\ref{lxlbol}). 
Only 20-30\% of our upper limit ratios lie at or below  log~$L_{\rm X}/L_{\rm bol}\sim -7$. 
Examination of our upper limit calculations for a 1 MK plasma reveal that the upper limits on X-ray luminosities are more strongly correlated with $N_H$ than with distance. 
The distribution of nondetections skews towards higher $N_H$ values compared to the detected CSPNe. 
It is furthermore apparent that the upper limits on X-ray luminosities for a 1 MK plasma have steadily worsened since the launch of {\it Chandra} due to its decreased soft X-ray sensitivity \citep{2004SPIE.5165..497M}. 
This combination of high $N_H$ and decreasing soft X-ray sensitivity likely explains the decrease the detection rate in later cycles. 

On the other hand, the upper limits calculated by assuming a higher temperature plasma (8 MK) are less sensitive to these effects and exhibit relatively stable behavior since the launch of {\it Chandra}. 
These ``hard'' X-ray emission upper limits can be used to further exclude potential spun-up binary companions. 
In this context, we briefly discuss the known close binary systems that were not detected in our observations. 

Five of the nine close binary systems in our sample are detected, and these CSPNe display some of the hardest X-ray spectra. 
The four close binary CSPNe that were not detected are Abell 65, Hen 2-11, NGC 2346, and NGC 6026, which have orbital periods of 1, 0.6, 16, and 0.53 days, respectively \citep{2009PASP..121..316D,2014A&A...562A..89J}.
\citet{2010AJ....140..319H} suggests the secondary star in NGC 6026 might be a compact white dwarf based on its compact size and high luminosity of $\sim10^{2}L_{\odot}$; hence this CSPN companion is unlikely to harbor a rejuvenated corona that would appear as a luminous X-ray source, and we do not consider it further. 
The heavily absorbed nucleus and uncertainty of the binary system parameters of the CSPN in NGC 2346 --- may consist of a compact sdO star and an A star, similar to NGC 1514 \citep{1981ApJ...250..240M} ---renders conclusions difficult. 
The upper limit X-ray luminosities derived from our observations of the remaining undetected binary CSPNe (Abell 65 and Hen 2-11) for an 8 MK plasma are 1.4 and 8.9$\times10^{29}~{\rm erg~s}^{-1}$, respectively (see Table~\ref{upperlimits}). 
For the secondaries found in Abell 65 \citep{2009AstBu..64..349S} and Hen 2-11 \citep{2014A&A...562A..89J}, the limit on log~$L_{\rm X}/L_{\rm bol,2}$ is -3.6 and -2.8, respectively. 
These two limits are similar to what we expect to observe from spun-up companions, thus deeper observations may yet detect saturated (or somewhat less than saturated) X-ray emission from the companions.

\section{Summary and Conclusions}

We have studied X-ray emission in the 0.3-3.0 keV energy band from the CSPNe observed thus far as part of the ChanPlaNS project. 
We performed spectral analysis of 20 point-like sources of CSPN X-ray emission and determined upper limits for 39 undetected CSPNe. 
The majority of the detected CSPNe spectra are well modeled as absorbed optically-thin thermal plasmas. 
The collective characteristics of the CSPNe and their X-ray emission suggest to two likely origins, (1) coronal activity associated with late-type binary companions and (2) self-shocking stellar winds.
These two processes can be distinguished based on CSPN X-ray emission characteristics. 
We find that the ratio of X-ray to bolometric luminosity, $L_{\rm X}/L_{\rm bol}$, and characteristic X-ray plasma temperature, $T_X$, can be used as first-order discriminators of binary companion versus shock origins for the X-ray emission: high-temperature plasmas ($T_{\rm X} \stackrel{>}{\sim} 6$ MK) 
with no clear correlation between $L_{\rm X}$ and the central star bolometric luminosity are likely due to active binary companions, whereas low temperature plasmas and values of log~$L_{\rm X}/L_{\rm bol}\sim-7$ are indicative of wind shocks.
For the hottest central stars ($\stackrel{>}{\sim} 130$ kK), we cannot rule out a contribution from photospheric emission in the spectral energy distribution for photon energies $\stackrel{>}{\sim}0.3$~keV, but only 2-3 CSPNe with X-ray emission fall into this range thus far.
Optical and/or infrared spectroscopic evidence of CSPN binarity and stellar winds, where available, generally provides support for the preceding wind and active late-type companion mechanisms as likely origins of CSPN X-ray emission. 

These two classes of CSPN X-ray emission -- self-shocking winds and binary companions -- may offer constraints for models of PN formation and shaping \citep[see reviews in][]{2002ARA&A..40..439B,2009PASP..121..316D}. 
Ejection of a common envelope, resulting in a close binary PN nucleus, is invoked as a mechanism to explain PN formation \citep{2009PASP..121..316D}. 
Such systems produce characteristic shapes and structures in their nebulae \citep{2009A&A...505..249M} and, based on the analysis of ChanPlaNS observations presented here, it appears hard X-ray emission may be characteristic of the resulting CSPN.
X-ray emission from self-shocking stellar winds provide unique constraints of the circumstellar conditions (densities, energetics, and radiation) of CSPNe.
The presence of such shocks also emphasize the instabilities of CSPN winds, providing a vital new ingredient for models that invoke shaping by a fast stellar wind. 
Meanwhile, alternative origins of the X-ray emission hint at polar outflows and equatorial material, structures often invoked to explain the shapes of PN via hydrodynamic interactions. 
By better establishing the potential origin(s) of the CSPN X-ray emission -- via observations of additional PNe selected so as to maximize the likelihood of detecting X-rays from binary companions and wind shocks, as well as through deeper observations of the brightest CSPN X-ray sources detected thus far -- we can further tap the potential of CSPN X-ray sources to constrain models describing PN shaping. 

\acknowledgements 

This research was supported via award numbers  GO1--12025A and GO3--14019A to Rochester Institute of Technology and GO3--14019B to Vanderbilt University issued by the Chandra X-ray Observatory Center, which is operated by the Smithsonian Astrophysical Observatory for and on behalf of NASA under contract NAS8-03060 (RIT).
R.S.'s contribution to the research described here was carried out at the Jet Propulsion Laboratory, California Institute of Technology, under a contract with NASA, and supported via an award issued by the Chandra X-Ray Observatory Center.
This research has made use of data obtained from the Chandra Data Archive and the Chandra Source Catalog, and software provided by the Chandra X-ray Center (CXC) in the application packages CIAO. 
This research has made use of data and/or software provided by the High Energy Astrophysics Science Archive Research Center (HEASARC), which is a service of the Astrophysics Science Division at NASA/GSFC and the High Energy Astrophysics Division of the Smithsonian Astrophysical Observatory.

\appendix 

\section{Photospheric Emission for Photon Energies $\geq0.3$~keV} \label{appendixA}

\subsection{Detection Threshold} 

We consider the potential for photospheric emission, modeled by blackbody distributions or NLTE atmosphere models, to emit X-ray photons in the 0.3-8.0 keV energy range.  
We begin by considering the best case scenario (appropriate for, e.g., NGC 7293 or M 27), i.e., a CSPN at a distance of 0.25 kpc from the Sun with negligible absorption.
Based on the typical background rates of our sample, we take the detection threshold of an on-axis observation on the Chandra ACIS-S3 chip in the 0.3-8.0 keV energy band as $F_X = 10^{-15} {\rm ~erg~cm}^{-2} {\rm ~s}^{-1}$. 
This corresponds to an X-ray luminosity threshold of 
\begin{equation} 
L_X \sim 8 \times 10^{27} {\rm ~erg~s}^{-1} \left( \frac{F_X}{10^{-15} {\rm ~erg~cm}^{-2} {\rm ~s}^{-1}} \right) \left( \frac{D}{0.25 {\rm ~kpc}} \right)^{2}. 
\end{equation}
For comparison with observations, we scale the X-ray luminosity by the central star bolometric luminosity, $L_{\rm bol}$, which can range from a few tens of $L_{\odot}$ up to $\sim10^{4}~L_{\odot}$. For a bright central star with $L_{\rm bol} = 10^{3} L_{\odot}$, log~$L_X/L_{\rm bol} \sim -8.7$, this suggests that to produce X-ray flux in the 0.3-8.0 keV band, log~$L_X/L_{\rm bol}$ must satisfy the following condition: 
\begin{equation}\label{limit}
{\rm log~}\frac{L_X}{L_{\rm bol}} \stackrel{>}{\sim} -8.7 + 2 {\rm ~log}\left(\frac{D}{0.25 {\rm ~kpc}}\right) - {\rm log}\left(\frac{L_{\rm bol}}{10^{3} L_{\odot}}\right).
\end{equation} 
In the following section we use the above criteria to determine the photospheric models that might contribute detectable X-ray flux for energies larger than 0.3 keV. 

\subsection{Photospheric Models} 

We use blackbody distributions and NLTE model atmospheres to model the photospheric emission of the CSPNe.
The blackbody model is determined by the temperature and model normalization, which is derived from the distance-scaled luminosity. 
The NLTE atmosphere models are characterized by temperature, model normalization, which is also derived from the distance-scaled luminosity, and a limited grid of log~$g$. 
Furthermore, we consider two chemically-distinct NLTE atmosphere models, an H-rich model with solar metal abundances and an H-poor model with metal abundances typical of PG1159-type CSPNe. 
For more information on these NLTE models, the reader is referred to \citet{2003A&A...403..709R}. 
These models are folded with the instrumental responses to produce observed X-ray spectra from which we calculate observed fluxes and median energies. 

\begin{figure*}[hb!]
\centering
\includegraphics[scale=0.5]{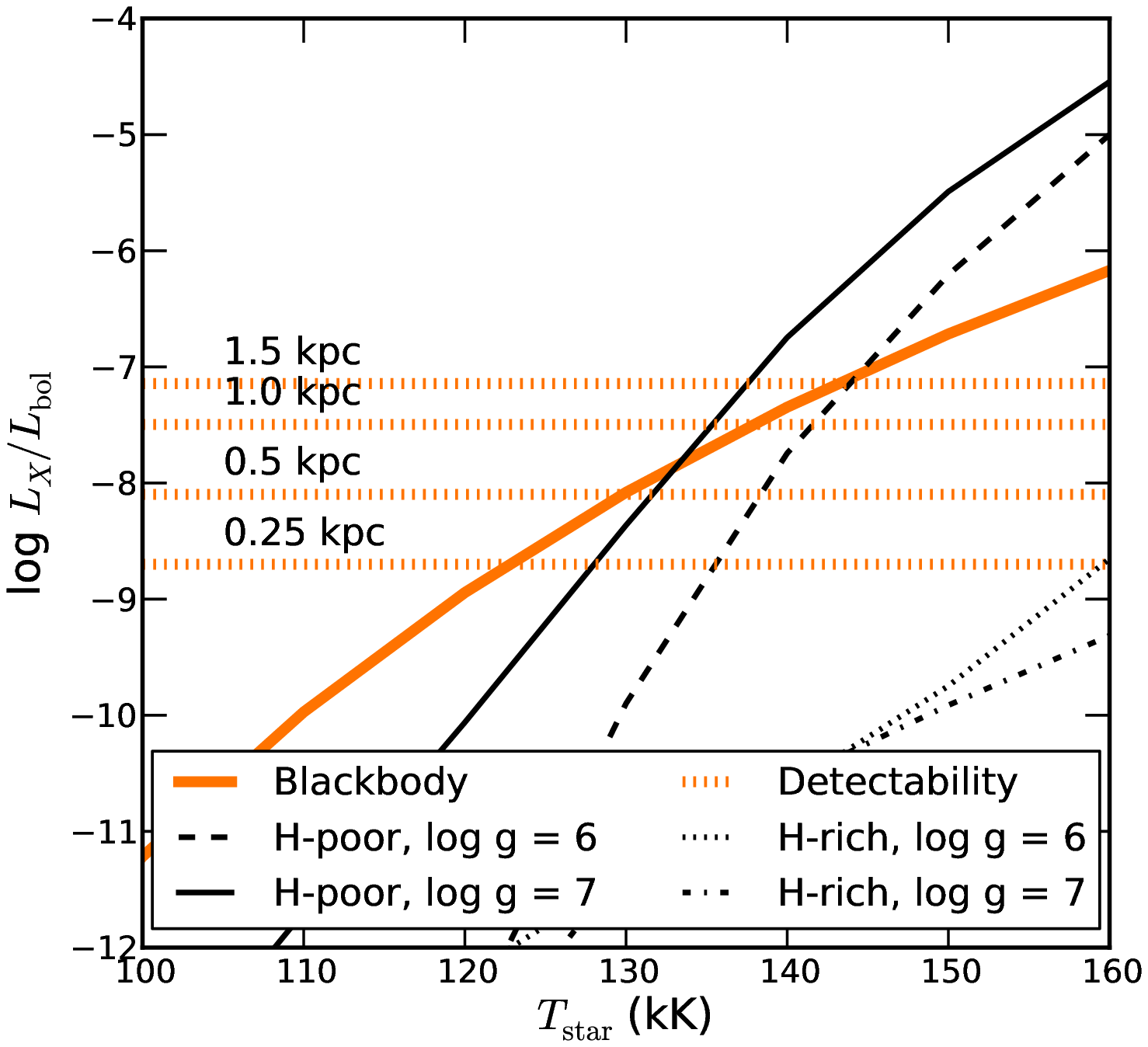}
\caption{$L_X/L_{\rm bol}$ ratios for photospheric models indicated in the legend and discussed in the text. Absorption-free distance limits using Equation~\ref{limit} are also indicated.
\label{appendix_fig1} }
\end{figure*}

In Figure~\ref{appendix_fig1}, we present the ratios of the X-ray luminosity in the 0.3-8.0 keV energy band with the bolometric luminosity of the central star for each model. 
The absorption-free detection limits for a range of distances of  ChanPlaNS targets are included in Figure~\ref{appendix_fig1}. 
These limits suggest that only blackbody distributions with $T_{\rm eff} > 120$ kK can produce detectable flux in the 0.3-8.0 keV energy band. 
However, at such high temperatures, NLTE effects like metal line-blanketing begin to dominate the high-energy, low-wavelength spectrum. 
The H-rich NLTE models are not capable of producing detected X-ray emission in the 0.3-8.0 keV energy range for the typical range of central star temperatures. 
However, H-poor, PG1159-type NLTE models with central star temperature ($\stackrel{>}{\sim} 130$ kK) can produce detectable X-ray flux in the 0.3-8.0 keV energy band, but the detectability decreases with distance and absorption. 

\begin{figure*}
\centering
\includegraphics[scale=0.5]{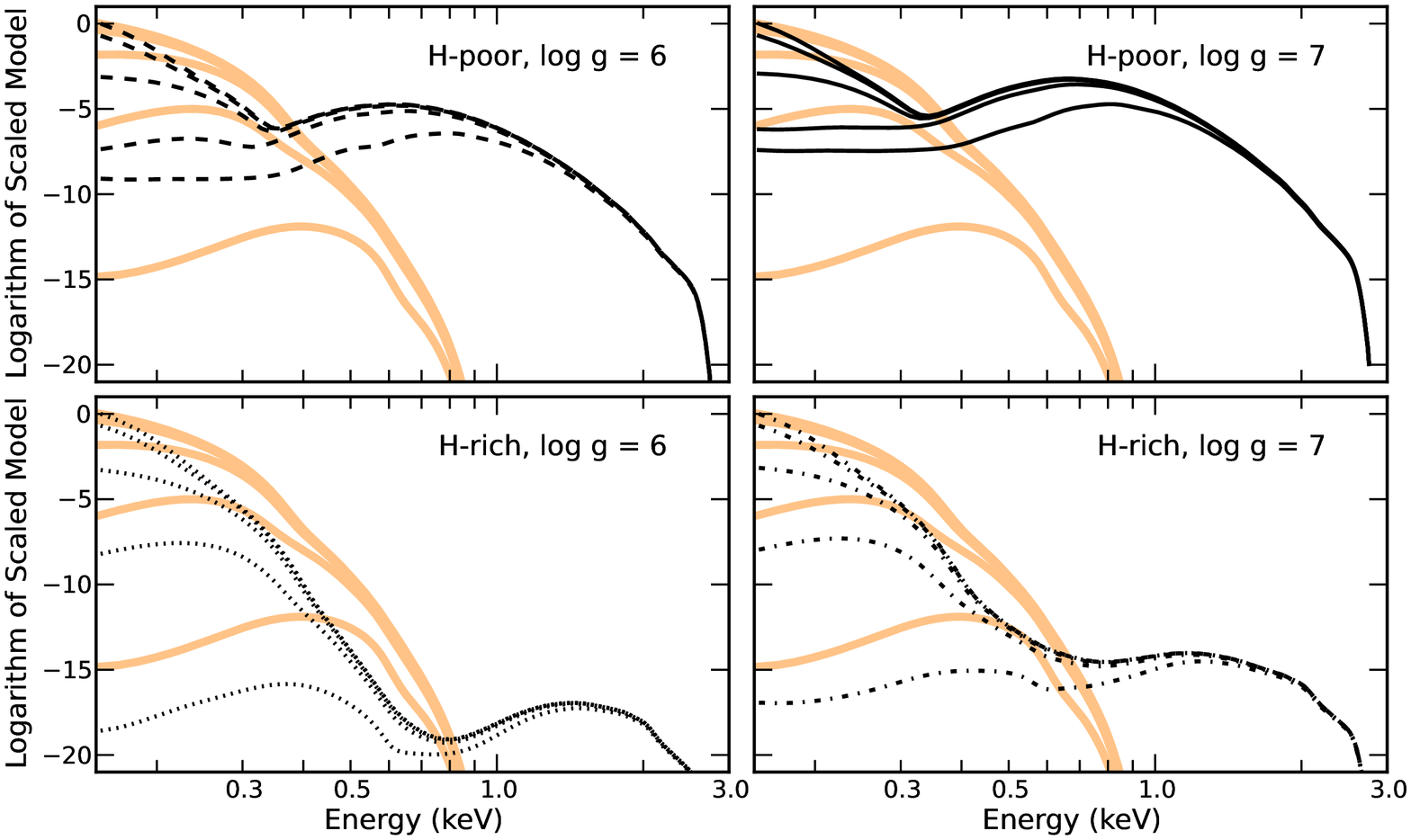}
\caption{X-ray spectra of the photospheric models for a central star temperature of 140 kK and log $N_H$ (cm$^{-2}$) of  19, 19.72, 20.44, 21.16, and 21.88. In each panel, the thick orange lines indicate a blackbody distribution while the black lines indicate various NLTE model atmospheres.  The models have been normalized by dividing by the maximum value of the model with log $N_H$ = 19.  These X-ray spectra suggest that H-deficient NLTE model atmospheres are capable of producing an excess X-ray flux above 0.5 keV.  
\label{appendix_fig2} }
\end{figure*}

\subsection{Detectable X-ray Flux from only the Hottest and Brightest H-deficient CSPNe}

\begin{figure*}
\centering
\includegraphics[scale=0.5]{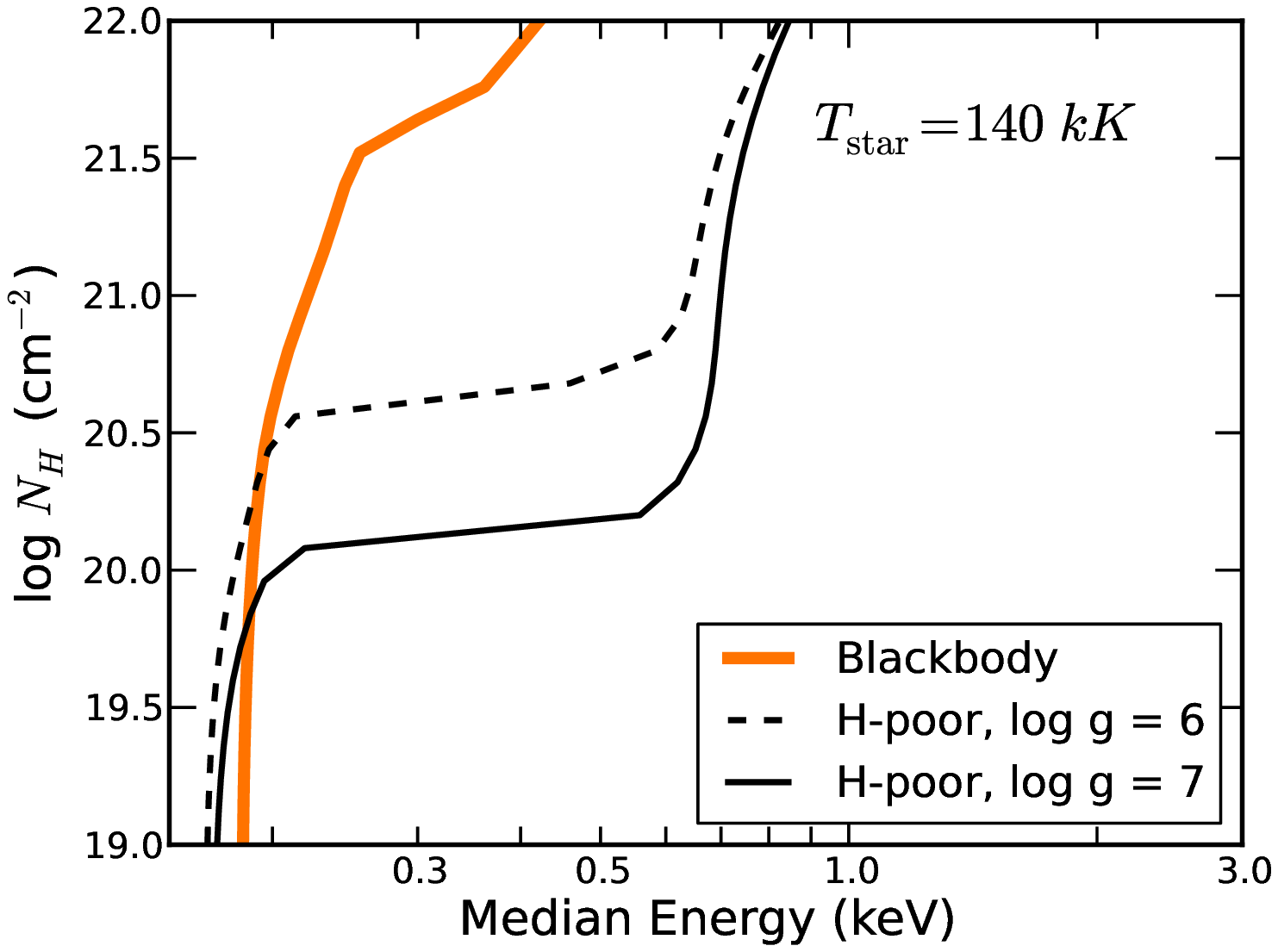}
\caption{The median energy and $N_H$ loci for blackbody and H-poor NLTE models for a central star temperature of 140 kK.
\label{appendix_fig3} }
\end{figure*}

Ultimately, these considerations suggest that only the hottest and brightest (nearest) H-deficient central stars are capable of producing detected X-ray flux in the 0.3-8.0 keV energy range. 
One such system, the central star of PN K 1-16, was considered in \citet{2013ApJ...766...26M}, where they found the NLTE model atmosphere successfully modeled soft X-ray emission below 0.3 keV and some X-ray emission above 0.3 keV, but not the excess flux at $\sim$0.35 keV. 
The authors modeled this excess flux with a carbon-rich plasma likely due to shocks in the circumstellar environment. 
Indeed, examining the X-ray spectra of the NLTE model atmospheres ($T_{\rm star} = 140$ kK provided in Figure~\ref{appendix_fig2}) reveals the excess is above 0.5 keV. 
The transition from low-energy median energies to that above 0.5 keV, is quite sudden as a function of $N_H$, as indicated in Figure~\ref{appendix_fig3}. 
This is due to the severe effect of absorption on the soft X-ray photons predicted by the NLTE models. 
Both \citet{2011MNRAS.417.2440H} and \citet{2013arXiv1307.2948K} include non-photospheric X-ray emitting components in their modeling of CSPNe spectra, indicating either NLTE models are not accurately predicting the X-ray emission or pointing towards a non-photospheric source of X-ray emission.

\begin{thebibliography}{}
\bibitem[Amnuel et al.(1985)]{1985Ap&SS.113...59A} Amnuel, P.~R., Guseinov, O.~K., Novruzova, K.~I., \& Rustamov, I.~S.\ 1985, \apss, 113, 59 
\bibitem[Anders \& Grevesse(1989)]{1989GeCoA..53..197A} Anders, E., \& Grevesse, N.\ 1989, \gca, 53, 197 
\bibitem[Anderson et al.(2011)]{2011ASPC..442..139A} Anderson, C.~S., Mossman, A.~E., Kim, D.-W., et al.\ 2011, Astronomical Data Analysis Software and Systems XX, 442, 139 
\bibitem[Arnaud(1996)]{1996ASPC..101...17A} Arnaud, K.~A.\ 1996, Astronomical Data Analysis Software and Systems V, 101, 17 
\bibitem[Balick \& Frank(2002)]{2002ARA&A..40..439B} Balick, B., \& Frank, A.\ 2002, \araa, 40, 439 
\bibitem[Bil{\'{\i}}kov{\'a} et al.(2010)]{2010AJ....140.1433B} Bil{\'{\i}}kov{\'a}, J., Chu, Y.-H., Gruendl, R.~A., \& Maddox, L.~A.\ 2010, \aj, 140, 1433 
\bibitem[Blackman et al.(2001a)]{2001Natur.409..485B} Blackman, E.~G., Frank, A., Markiel, J.~A., Thomas, J.~H., \& Van Horn, H.~M.\ 2001, \nat, 409, 485 
\bibitem[Blackman et al.(2001b)]{2001ApJ...546..288B} Blackman, E.~G., Frank, A., \& Welch, C.\ 2001, \apj, 546, 288 
\bibitem[Bloecker(1995)]{1995A&A...299..755B} Bloecker, T.\ 1995, \aap, 299, 755 
\bibitem[Bohlin et al.(1978)]{1978ApJ...224..132B} Bohlin, R.~C., Savage, B.~D., \& Drake, J.~F.\ 1978, \apj, 224, 132 
\bibitem[Bonito et al.(2011)]{2011ApJ...737...54B} Bonito, R., Orlando, S., Miceli, M., et al.\ 2011, \apj, 737, 54 
\bibitem[Cahn et al.(1992)]{1992A&AS...94..399C} Cahn, J.~H., Kaler, J.~B., \& Stanghellini, L.\ 1992, \aaps, 94, 399 
\bibitem[Cahn(1984)]{1984ApJ...279..304C} Cahn, J.~H.\ 1984, \apj, 279, 304 
\bibitem[Carter et al.(2003)]{2003ASPC..295..477C} Carter, C., Karovska, M., Jerius, D., Glotfelty, K., \& Beikman, S.\ 2003, Astronomical Data Analysis Software and Systems XII, 295, 477 
\bibitem[Cassinelli et al.(1981)]{1981ApJ...250..677C} Cassinelli, J.~P., Waldron, W.~L., Sanders, W.~T., et al.\ 1981, \apj, 250, 677 
\bibitem[Cerruti-Sola \& Perinotto(1985)]{1985ApJ...291..237C} Cerruti-Sola, M., \& Perinotto, M.\ 1985, \apj, 291, 237 
\bibitem[Chu et al.(2007)]{2007ASPC..372..337C} Chu, Y.-H., Gruendl, R.~A., Guerrero, M.~A., \& Su, K.~Y.-L.\ 2007, 15th European Workshop on White Dwarfs, 372, 337 
\bibitem[Chu et al.(2004)]{2004AJ....127..477C} Chu, Y.-H., Guerrero, M.~A., Gruendl, R.~A., \& Webbink, R.~F.\ 2004, \aj, 127, 477 
\bibitem[Chu et al.(2001)]{2001ApJ...553L..69C} Chu, Y.-H., Guerrero, M.~A., Gruendl, R.~A., Williams, R.~M., \& Kaler, J.~B.\ 2001, \apjl, 553, L69 
\bibitem[Churazov et al.(1996)]{1996ApJ...471..673C} Churazov, E., Gilfanov, M., Forman, W., \& Jones, C.\ 1996, \apj, 471, 673 
\bibitem[Ciardullo et al.(1999)]{1999AJ....118..488C} Ciardullo, R., Bond, H.~E., Sipior, M.~S., et al.\ 1999, \aj, 118, 488 
\bibitem[Cohen et al.(2014)]{2014MNRAS.439..908C} Cohen, D.~H., Wollman, E.~E., Leutenegger, M.~A., et al.\ 2014, \mnras, 439, 908 
\bibitem[Danehkar et al.(2012)]{2012IAUS..282..470D} Danehkar, A., Frew, D.~J., Parker, Q.~A., \& De Marco, O.\ 2012, IAU Symposium, 282, 470 
\bibitem[De Marco 
\& Soker(2002)]{2002PASP..114..602D} De Marco, O., \& Soker, N.\ 2002, \pasp, 114, 602 
\bibitem[De Marco(2009)]{2009PASP..121..316D} De Marco, O.\ 2009, \pasp, 121, 316 
\bibitem[Feibelman(2000)]{2000ApJ...542..957F} Feibelman, W.~A.\ 2000, \apj, 542, 957 
\bibitem[Feibelman(1997)]{1997ApJS..112..193F} Feibelman, W.~A.\ 1997, \apjs, 112, 193 
\bibitem[Feldmeier et al.(1997)]{1997A&A...322..878F} Feldmeier, A., Puls, J., \& Pauldrach, A.~W.~A.\ 1997, \aap, 322, 878 
\bibitem[Foster et al.(2012)]{2012ApJ...756..128F} Foster, A.~R., Ji, L., Smith, R.~K., \& Brickhouse, N.~S.\ 2012, \apj, 756, 128 
\bibitem[Frank et al.(2002)]{2002apa..book.....F} Frank, J., King, A., \& Raine, D.~J.\ 2002, Accretion Power in Astrophysics, by Juhan Frank and Andrew King and Derek Raine, pp.~398.~ISBN 0521620538.~Cambridge, UK: Cambridge University Press, February 2002.
\bibitem[Frankowski \& Soker(2009)]{2009NewA...14..654F} Frankowski, A., \& Soker, N.\ 2009, New Astronomy, 14, 654 
\bibitem[Paper(II)]{ChanPlaNSII} Freeman et al., Paper II, in prep
\bibitem[Frew et al.(2013)]{2013MNRAS.431....2F} Frew, D.~J., Boji{\v c}i{\'c}, I.~S., \& Parker, Q.~A.\ 2013, \mnras, 431, 2 
\bibitem[Frew(2008)]{2008PhDT.......109F} Frew, D.~J.\ 2008, Ph.D.~Thesis,  
\bibitem[Fruscione et al.(2006)]{2006SPIE.6270E..60F} Fruscione, A., et  al.\ 2006, \procspie, 6270
\bibitem[G{\"u}del(2004)]{2004A&ARv..12...71G} G{\"u}del, M.\ 2004, \aapr, 12, 71 
\bibitem[G{\"u}del \& Naz{\'e}(2009)]{2009A&ARv..17..309G} G{\"u}del, M., \& Naz{\'e}, Y.\ 2009, \aapr, 17, 309 
\bibitem[Gagn{\'e} et al.(2005)]{2005ApJ...628..986G} Gagn{\'e}, M., Oksala, M.~E., Cohen, D.~H., et al.\ 2005, \apj, 628, 986 
\bibitem[Getman et al.(2010)]{2010ApJ...708.1760G} Getman, K.~V., Feigelson, E.~D., Broos, P.~S., Townsley, L.~K., \& Garmire, G.~P.\ 2010, \apj, 708, 1760 
\bibitem[Goldman et al.(2004)]{2004AJ....128.1711G} Goldman, D.~B., Guerrero, M.~A., Chu, Y.-H., \& Gruendl, R.~A.\ 2004, \aj, 128, 1711 
\bibitem[Gondoin(2007)]{2007A&A...464.1101G} Gondoin, P.\ 2007, \aap, 464, 1101 
\bibitem[Gregory \& Loredo(1992)]{1992ApJ...398..146G} Gregory, P.~C., \& Loredo, T.~J.\ 1992, \apj, 398, 146
\bibitem[Grimm et al.(2009)]{2009ApJ...690..128G} Grimm, H.-J., McDowell, J., Fabbiano, G., \& Elvis, M.\ 2009, \apj, 690, 128 
\bibitem[Gruendl et al.(2001)]{2001AJ....122..308G} Gruendl, R.~A., Chu, Y.-H., O'Dwyer, I.~J., \& Guerrero, M.~A.\ 2001, \aj, 122, 308 
\bibitem[Guerrero et al.(2005)]{2005A&A...430L..69G} Guerrero, M.~A., Chu, Y.-H., Gruendl, R.~A., \& Meixner, M.\ 2005, \aap, 430, L69 
\bibitem[Guerrero \& De Marco(2013)]{2013A&A...553A.126G} Guerrero, M.~A., \& De Marco, O.\ 2013, \aap, 553, A126 
\bibitem[Guerrero et al.(2012)]{2012ApJ...755..129G} Guerrero, M.~A., Ruiz, N., Hamann, W.-R., et al.\ 2012, \apj, 755, 129 
\bibitem[Guerrero et al.(2000)]{2000ApJS..129..295G} Guerrero, M.~A., Chu, Y.-H., \& Gruendl, R.~A.\ 2000, \apjs, 129, 295 
\bibitem[Guerrero et al.(2001)]{2001ApJ...553L..55G} Guerrero, M.~A., Chu, Y.-H., Gruendl, R.~A., Williams, R.~M., \& Kaler, J.~B.\ 2001, \apjl, 553, L55 
\bibitem[Herald \& Bianchi(2011)]{2011MNRAS.417.2440H} Herald, J.~E., \& Bianchi, L.\ 2011, \mnras, 417, 2440 
\bibitem[Hillwig et al.(2010)]{2010AJ....140..319H} Hillwig, T.~C., Bond, H.~E., Af{\c s}ar, M., \& De Marco, O.\ 2010, \aj, 140, 319 
\bibitem[Hoogerwerf et al.(2007)]{2007ApJ...670..442H} Hoogerwerf, R., Szentgyorgyi, A., Raymond, J., et al.\ 2007, \apj, 670, 442 
\bibitem[Jacob et al.(2013)]{2013A&A...558A..78J} Jacob, R., Sch{\"o}nberner, D., \& Steffen, M.\ 2013, \aap, 558, A78 
\bibitem[Jones et al.(2014)]{2014A&A...562A..89J} Jones, D., Boffin, H.~M.~J., Miszalski, B., et al.\ 2014, \aap, 562, A89 
\bibitem[Kaler \& Lutz(1985)]{1985PASP...97..700K} Kaler, J.~B., \& Lutz, J.~H.\ 1985, \pasp, 97, 700 
\bibitem[Kaler \& Shaw(1984)]{1984ApJ...278..195K} Kaler, J.~B., \& Shaw, R.~A.\ 1984, \apj, 278, 195 
\bibitem[Kaschinski et al.(2013)]{2013arXiv1307.2948K} Kaschinski, C.~B., Pauldrach, A.~W.~A., \& Hoffmann, T.~L.\ 2013, arXiv:1307.2948 
\bibitem[Kastner et al.(2000)]{2000ApJ...545L..57K} Kastner, J.~H., Soker, N., Vrtilek, S.~D., \& Dgani, R.\ 2000, \apjl, 545, L57 
\bibitem[Kastner et al.(2005)]{2005ApJS..160..511K} Kastner, J.~H., Franz, G., Grosso, N., et al.\ 2005, \apjs, 160, 511 
\bibitem[Paper(I)]{2012AJ....144...58K} Kastner, J.~H., Montez, R., Jr., Balick, B., et al.\ 2012, \aj, 144, 58 
\bibitem[Kwok et al.(1978)]{1978ApJ...219L.125K} Kwok, S., Purton, C.~R., \& Fitzgerald, P.~M.\ 1978, \apjl, 219, L125 
\bibitem[Leone et al.(2014)]{2014A&A...563A..43L} Leone, F., Corradi, R.~L.~M., Mart{\'{\i}}nez Gonz{\'a}lez, M.~J., Asensio Ramos, A., \& Manso Sainz, R.\ 2014, \aap, 563, A43 
\bibitem[Li et al.(2004)]{2004ApJ...610.1204L} Li, J., Kastner, J.~H., Prigozhin, G.~Y., et al.\ 2004, \apj, 610, 1204 
\bibitem[Li et al.(2003)]{2003ApJ...590..586L} Li, J., Kastner, J.~H., Prigozhin, G.~Y., \& Schulz, N.~S.\ 2003, \apj, 590, 586 
\bibitem[Marshall et al.(2004)]{2004SPIE.5165..497M} Marshall, H.~L., Tennant, A., Grant, C.~E., et al.\ 2004, \procspie, 5165, 497 
\bibitem[Mendez(1991)]{1991IAUS..145..375M} Mendez, R.~H.\ 1991, Evolution of Stars: the Photospheric Abundance Connection, 145, 375 
\bibitem[Mendez \& Niemela(1981)]{1981ApJ...250..240M} Mendez, R.~H., \& Niemela, V.~S.\ 1981, \apj, 250, 240 
\bibitem[Miszalski et al.(2009)]{2009A&A...505..249M} Miszalski, B., Acker, A., Parker, Q.~A., \& Moffat, A.~F.~J.\ 2009, \aap, 505, 249 
\bibitem[Modigliani et al.(1993)]{1993ApJ...415..258M} Modigliani, A., Patriarchi, P., \& Perinotto, M.\ 1993, \apj, 415, 258 
\bibitem[Montez et al.(2010)]{2010ApJ...721.1820M} Montez, R., Jr., De Marco, O., Kastner, J.~H., \& Chu, Y.-H.\ 2010, \apj, 721, 1820 
\bibitem[Montez \& Kastner(2013)]{2013ApJ...766...26M} Montez, R., Jr., \& Kastner, J.~H.\ 2013, \apj, 766, 26 
\bibitem[Morrison \& McCammon(1983)]{1983ApJ...270..119M} Morrison, R., \& McCammon, D.\ 1983, \apj, 270, 119 
\bibitem[Motch et al.(1993)]{1993A&A...268..561M} Motch, C., Werner, K., \& Pakull, M.~W.\ 1993, \aap, 268, 561 
\bibitem[Muerset et al.(1997)]{1997A&A...319..201M} Muerset, U., Wolff, B., \& Jordan, S.\ 1997, \aap, 319, 201 
\bibitem[Napiwotzki \& Schoenberner(1991)]{1991A&A...249L..16N} Napiwotzki, R., \& Schoenberner, D.\ 1991, \aap, 249, L16 
\bibitem[Napiwotzki \& Schoenberner(1995)]{1995A&A...301..545N} Napiwotzki, R., \& Schoenberner, D.\ 1995, \aap, 301, 545 
\bibitem[Naz{\'e} et al.(2011)]{2011ApJS..194....7N} Naz{\'e}, Y., Broos, P.~S., Oskinova, L., et al.\ 2011, \apjs, 194, 7 
\bibitem[O'Dwyer et al.(2003)]{2003AJ....125.2239O} O'Dwyer, I.~J., Chu, Y.-H., Gruendl, R.~A., Guerrero, M.~A., \& Webbink, R.~F.\ 2003, \aj, 125, 2239 
\bibitem[Pandel et al.(2005)]{2005ApJ...626..396P} Pandel, D., C{\'o}rdova, F.~A., Mason, K.~O., \& Priedhorsky, W.~C.\ 2005, \apj, 626, 396 
\bibitem[Patriarchi \& Perinotto(1997)]{1997A&AS..126..385P} Patriarchi, P., \& Perinotto, M.\ 1997, \aaps, 126, 385 
\bibitem[Patriarchi \& Perinotto(1995)]{1995A&AS..110..353P} Patriarchi, P., \& Perinotto, M.\ 1995, \aaps, 110, 353 
\bibitem[Patriarchi \& Perinotto(1991)]{1991A&AS...91..325P} Patriarchi, P., \& Perinotto, M.\ 1991, \aaps, 91, 325 
\bibitem[Patriarchi \& Perinotto(1991)]{1991PASAu...9..309P} Patriarchi, P., \& Perinotto, M.\ 1991, Proceedings of the Astronomical Society of Australia, 9, 309
\bibitem[Pauldrach et al.(2004)]{2004A&A...419.1111P} Pauldrach, A.~W.~A., Hoffmann, T.~L., \& M{\'e}ndez, R.~H.\ 2004, \aap, 419, 1111 
\bibitem[Perinotto et al.(2004)]{2004A&A...414..993P} Perinotto, M., Sch{\"o}nberner, D., Steffen, M., \& Calonaci, C.\ 2004, \aap, 414, 993 
\bibitem[Pizzolato et al.(2003)]{2003A&A...397..147P} Pizzolato, N., Maggio, A., Micela, G., Sciortino, S., \& Ventura, P.\ 2003, \aap, 397, 147 
\bibitem[Prinja(1990)]{1990A&A...232..119P} Prinja, R.~K.\ 1990, \aap, 232, 119 
\bibitem[Prinja et al.(2007)]{2007MNRAS.382..299P} Prinja, R.~K., Hodges, S.~E., Massa, D.~L., Fullerton, A.~W., \& Burnley, A.~W.\ 2007, \mnras, 382, 299 
\bibitem[Prinja et al.(2012)]{2012ApJ...759L..28P} Prinja, R.~K., Massa, D.~L., \& Cantiello, M.\ 2012, \apjl, 759, L28 
\bibitem[Prinja et al.(1990)]{1990ApJ...361..607P} Prinja, R.~K., Barlow, M.~J., \& Howarth, I.~D.\ 1990, \apj, 361, 607 
\bibitem[Prinja \& Urbaneja(2014)]{2014MNRAS.440.2684P} Prinja, R.~K., \& Urbaneja, M.~A.\ 2014, \mnras, 440, 2684 
\bibitem[Ramstedt et al.(2012)]{2012A&A...543A.147R} Ramstedt, S., Montez, R., Kastner, J., \& Vlemmings, W.~H.~T.\ 2012, \aap, 543, A147 
\bibitem[Rauch(2003)]{2003A&A...403..709R} Rauch, T.\ 2003, \aap, 403, 709 
\bibitem[Ressler et al.(2010)]{2010AJ....140.1882R} Ressler, M.~E., Cohen, M., Wachter, S., et al.\ 2010, \aj, 140, 1882 
\bibitem[Ruiz et al.(2013)]{2013ApJ...767...35R} Ruiz, N., Chu, Y.-H., Gruendl, R.~A., et al.\ 2013, \apj, 767, 35 
\bibitem[Schmidt et al.(2003)]{2003ApJ...595.1101S} Schmidt, G.~D., Harris, H.~C., Liebert, J., et al.\ 2003, \apj, 595, 1101 
\bibitem[Shimansky et al.(2009)]{2009AstBu..64..349S} Shimansky, V.~V., Pozdnyakova, S.~A., Borisov, N.~V., et al.\ 2009, Astrophysical Bulletin, 64, 349 
\bibitem[Schoenberner(1983)]{1983ApJ...272..708S} Schoenberner, D.\ 1983, \apj, 272, 708 
\bibitem[Soker(2001)]{2001MNRAS.328.1081S} Soker, N.\ 2001, \mnras, 328, 1081 
\bibitem[Soker \& Kastner(2002)]{2002ApJ...570..245S} Soker, N., \& Kastner, J.~H.\ 2002, \apj, 570, 245 
\bibitem[Su et al.(2007)]{2007ApJ...657L..41S} Su, K.~Y.~L., Chu, Y.-H., Rieke, G.~H., et al.\ 2007, \apjl, 657, L41 
\bibitem[Vanlandingham et al.(2005)]{2005AJ....130..734V} Vanlandingham, K.~M., Schmidt, G.~D., Eisenstein, D.~J., et al.\ 2005, \aj, 130, 734 
\bibitem[V{\'a}zquez et al.(1999)]{1999MNRAS.308..939V} V{\'a}zquez, R., L{\'o}pez, J.~A., Miranda, L.~F., Torrelles, J.~M., \& Meaburn, J.\ 1999, \mnras, 308, 939 
\bibitem[Walder(1998)]{1998Ap&SS.260..243W} Walder, R.\ 1998, \apss, 260, 243 
\bibitem[Weidmann \& Gamen(2011)]{2011A&A...526A...6W} Weidmann, W.~A., \& Gamen, R.\ 2011, \aap, 526, A6 
\bibitem[Werner \& Herwig(2006)]{2006PASP..118..183W} Werner, K., \& Herwig, F.\ 2006, \pasp, 118, 183 
\end{thebibliography}
\end{document}